# Fast interpolation of sparse multivariate polynomials[*][†]

Joris van der Hoeven[a], Grégoire Lecerf[b]

Laboratoire d'informatique de l'École polytechnique (LIX, UMR 7161)
CNRS, École polytechnique, Institut Polytechnique de Paris
Bâtiment Alan Turing, CS35003
1, rue Honoré d'Estienne d'Orves
91120 Palaiseau, France

a. *Email:* vdhoeven@lix.polytechnique.fr
b. *Email:* lecerf@lix.polytechnique.fr

Consider a sparse multivariate polynomial $f$ with integer coefficients. Assume that $f$ is represented as a "modular black box polynomial", e.g. via an algorithm to evaluate $f$ at arbitrary integer points, modulo arbitrary positive integers. The problem of sparse interpolation is to recover $f$ in its usual sparse representation, as a sum of coefficients times monomials. For the first time we present a quasi-optimal algorithm for this task.

## 1. Introduction

Consider a multivariate integer polynomial $f \in \mathbb{Z}[x_0, \ldots, x_{n-1}]$. Then $f$ can uniquely be written as a sum

$$f = c_0 x^{e_0} + \cdots + c_{t-1} x^{e_{t-1}}, \qquad (1.1)$$

where $c_0, \ldots, c_{t-1} \in \mathbb{Z}^{\neq} := \{i \in \mathbb{Z} : i \neq 0\}$ and $e_0, \ldots, e_{t-1} \in \mathbb{N}^n$ are pairwise distinct. Here we understand that $x^\alpha := x_0^{\alpha_0} \cdots x_{n-1}^{\alpha_{n-1}}$ for any $\alpha = (\alpha_0, \ldots, \alpha_{n-1}) \in \mathbb{N}^n$. We call (1.1) the *sparse representation* of $f$.

In this paper, we assume that $f$ is *not* explicitly given through its sparse representation and that we only have a program for evaluating $f$. The goal of *sparse interpolation* is to recover the sparse representation of $f$.

Theoretically speaking, we could simply evaluate $f$ at a single point $a = (a_0, \ldots, a_{n-1})$ with $a_0 = 2^E, a_1 = 2^{E^2}, \ldots, a_{n-1} = 2^{E^n}$ for some sufficiently large positive integer $E$. Then the sparse representation of $f$ can directly be read off from the binary digits of $f(a)$. However, the bit-complexity of this method is terrible, since the bit-size of $f(a)$ typically becomes huge.

---

[*]. This work has partly been supported by the French ANR-22-CE48-0016 NODE project.

[†]. This article has been written using GNU T<sub>E</sub>X<sub>MACS</sub> [23].

In order to get a better grip on the bit-complexity to evaluate and then interpolate $f$, we will assume that we actually have a program to evaluate $f$ modulo $m$ for any positive integer modulus $m$. We denote by $C_f(s)$ the cost to evaluate $f$ for a modulus with $m < 2^s$ and we assume that the *average cost per bit* $A_f(s) := C_f(s)/s$ is a non-decreasing function that grows not too fast as a function of $s$. Since the blackbox function should at least read its $n$ input values, we also assume that $n = O(A_f(s))$.

As in [15], for complexity estimates we use a *random memory access* (RAM) machine over a finite alphabet along with the *soft-Oh* notation: $f(z) = \tilde{O}(g(z))$ means that $f(z) = g(z) (\log (g(z)))^{O(1)}$. The machine is expected to have an instruction for generating a random bit in constant time (see section 2 for the precise computational model and hypotheses that we use). Assuming that we are given a bound $S$ for the bit-size of $f$, the main result of this paper is the following:

THEOREM 1.1. *There is a Monte Carlo algorithm that takes a modular blackbox representation of a sparse polynomial $f \in \mathbb{Z}[x_0, \ldots, x_{n-1}]$ of bit-size $\leqslant S$ as input and that interpolates f in time $A_f(S) \tilde{O}(S)$ with a probability of success at least $1/2$.*

The problem of sparse interpolation has a long history and goes back to work by Prony in the 18[th] century [40]. The first modern fast algorithm is due to Ben-Or and Tiwari [6]. Their work spawned a new area of research in computer algebra together with early implementations [11, 13, 19, 29, 32, 33, 36, 44]. We refer to [41] and [39, section 3] for nice surveys.

Modern research on sparse interpolation has developed in two directions. The first theoretical line of research has focused on rigorous and general complexity bounds [2, 3, 3, 4, 14, 17, 30]. The second direction concerns implementations, practical efficiency, and applications of sparse interpolation [5, 18, 20, 22, 25, 26, 28, 31, 34, 35].

The present paper mainly falls in the first line of research, although we will briefly discuss practical aspects in section 5. The proof of our main theorem relies on some number theoretical facts about prime numbers that will be recalled in section 2.3. There is actually a big discrepancy between empirical observations about prime numbers and hard theorems that one is able to prove. Because of this, our algorithms involve constant factors that are far more pessimistic than the ones that can be used in practice. Our algorithms also involve a few technical complications in order to cover all possible cases, including very large exponents that are unlikely to occur in practice.

Our paper borrows many techniques from [17, 39] that deal with the particular case when $f$ is a univariate polynomial. In principle, the multivariate case can be reduced to this case: setting $g(t) = f(t, t^E, \ldots, t^{E^{n-1}})$ for a sufficiently large $E \in \mathbb{N}$, the interpolation of $f$ reduces to the interpolation of $g$. However, this reduction is optimal only if the entries of the vector exponents $e_i \in \mathbb{N}^n$ are all approximately of the same bit-size. One interesting case that is not well covered by this reduction is when the number of variables $n$ is large and when the exponent vectors $e_i$ are themselves sparse in the sense that only a few entries are non-zero.

In order to cover sparse or unbalanced exponent vectors in a more efficient way, we will introduce a new technique in section 3. The idea is to compress such exponent vectors using random projections. With high probability, it will be possible to reconstruct the actual exponent vectors from their projections. We regard this section as the central technical contribution of our paper. Another difference with [17, 39] is that, in our main theorem, the coefficients $c_i$ of $f$ are allowed to wildly vary in bit-size. This is not as big a problem as the one of unbalanced exponent vectors, but still needs to be dealt with carefully, as we will do in section 4. Let us further mention that random projections of the exponents were previously implemented in [4] in order to reduce the multivariate case to the univariate one: monomial collisions were avoided in this way but the reconstruction of the exponents needed linear algebra and could not really catch sparse or unbalanced exponents. The proof of Theorem 1.1 will be completed at the end of section 4. Section 5 will address the practical aspects of our new method. An important inspiration behind the techniques from section 3 and its practical variants is the mystery ball game from [22]; this connection will also be discussed in section 5.

Our paper focuses on the case when $f$ has integer coefficients, but our algorithm can be easily adapted to rational coefficients as well, essentially by appealing to rational reconstruction [15, Chapter 5, section 5.10] during the proof of Lemma 4.8. However when $f$ has rational coefficients, its blackbox might include divisions and therefore raise "division by zero" errors occasionally. This makes the probability analysis and the worst case complexity bounds more difficult to analyze, so we preferred to postpone this study to another paper.

Our algorithm should also be applicable to various other coefficient rings of characteristic zero. However it remains an open challenge to develop similar algorithms for coefficient rings of small positive characteristic.

**Notation.** Throughout this paper, we will use the following notation:

$$\mathbb{N} := \{0, 1, 2, \ldots\}$$
$$\mathbb{N}^{>} := \{1, 2, 3, \ldots\}$$

For any $k \in \mathbb{N}$, we also define

$$\mathbb{N}_k := \{0, \ldots, k-1\}$$
$$\mathbb{Z}_k := \mathbb{N}_k - \left\lfloor \frac{k-1}{2} \right\rfloor.$$

We may use both $\mathbb{N}_k$ and $\mathbb{Z}_k$ as sets of canonical representatives modulo $k$. Given $i \in \mathbb{Z}$ and depending on the context, we write $i \operatorname{rem} k$ for the unique $r \in \mathbb{N}_k$ or $r \in \mathbb{Z}_k$ with $i - r \in k\mathbb{Z}$.

## 2. PRELIMINARIES

This section presents sparse data structures, computational models, and quantitative results about prime distributions. At the end, an elementary fast algorithm is presented for testing the divisibility of several integers by several prime numbers.

### 2.1. Sparse polynomials

We order $\mathbb{N}^n$ lexicographically by $<$. Given formal indeterminates $x_0, \ldots, x_{n-1}$ and an *exponent* $e = (e_0, \ldots, e_{n-1}) \in \mathbb{N}^n$, we define $x^e := x_0^{e_0} \cdots x_{n-1}^{e_{n-1}}$. We define the *bit-size* of an integer $i \in \mathbb{N}$ as $\sigma_i := \min\{s \in \mathbb{N} : i < 2^s\}$. In particular, $\sigma_0 = 0$, $\sigma_1 = 1$, $\sigma_2 = 2$, etc. We define the *bit-size* of an exponent tuple $e \in \mathbb{N}^n$ by $\sigma_e := e_0 + \cdots + e_{n-1}$. We extend these definitions to the cases when $i \in \mathbb{Z}$ and $e \in \mathbb{Z}^n$ by setting $\sigma_i := \sigma_{|i|}$ and $\sigma_e := \sigma_{|e|}$, where $|e| = (|e_0|, \ldots, |e_{n-1}|)$.

Now consider a multivariate polynomial $f \in \mathbb{Z}[x_0, \ldots, x_{n-1}]$. Then $f$ can uniquely be written as a sum

$$f = c_0 x^{e_0} + \cdots + c_{T-1} x^{e_{T-1}},$$

where $c_0, \ldots, c_{T-1} \in \mathbb{Z}^{\neq}$ and $e_0, \ldots, e_{T-1} \in \mathbb{N}^n$ are such that $e_0 < \cdots < e_{T-1}$. We call this the *sparse representation* of $f$. We call $e_0, \ldots, e_{T-1}$ the *exponents* of $f$ and $c_0, \ldots, c_{T-1}$ the corresponding *coefficients*. We also say that $c_0 x^{e_0}, \ldots, c_{T-1} x^{e_{T-1}}$ are the *terms* of $f$ and we call $\operatorname{supp} f := \{e_0, \ldots, e_{T-1}\}$ the *support* of $f$. Any non-zero $e_{i,j}$ with $i \in \mathbb{N}_T$ and $j \in \mathbb{N}_n$ is called an *individual exponent* of $f$. We define $\sigma_f := \sigma_{c_0} + \sigma_{e_0} + \cdots + \sigma_{c_{T-1}} + \sigma_{e_{T-1}}$ to be the *bit-size* of $f$.

**Remark 2.1.** For the complexity model we could have chosen a multi-tape Turing machine, but this would have led to more tedious cost analyses. In fact on a Turing tape, we would actually need to indicate the ends of numbers using adequate markers. Using a liberal notion of "bit", which allows for the storage of such markers in a single bit, the *Turing bit-size* of an integer $i \in \mathbb{N}$ then becomes $\sigma_i^* := \sigma_i + 1$. For $i \in \mathbb{N}^>$, we also define $\sigma_{-i}^* := \sigma_i^* + 1$. Exponents $e = (e_0, \ldots, e_{n-1})$ can be stored by appending the representations of $e_0, \ldots, e_{n-1}$, but this is suboptimal in the case when only a few entries of $e$ are non-zero. For such "sparse exponents", one prefers to store the pairs $(i, e_i)$ for which $e_i \neq 0$, again using suitable markers. For this reason, the *Turing bit-size* of $e$ becomes $\sigma_e^* := \min(\sigma_0^* + \cdots + \sigma_{n-1}^*, \sum_{e_i \neq 0} (\sigma_i^* + \sigma_{e_i}^*)) + 1$.

## 2.2. Modular blackbox polynomials

Throughout this paper, we will analyze bit complexities in the RAM model as in [15]. In this model, it is known [21] that two $n$-bit integers can be multiplied in time $O(n \log n)$. As a consequence, given an $n$-bit modulus $m \in \mathbb{N}^>$, the ring operations in $\mathbb{Z}/m\mathbb{Z}$ can be done with the same complexity [9, 15]. Inverses can be computed in time $O(n \log^2 n)$, whenever they exist [9, 15]. For randomized algorithms, we assume that we have an instruction for generating a random bit in constant time.

Consider a polynomial $f \in \mathbb{Z}[x_0, \ldots, x_{n-1}]$. A *modular blackbox representation* for $f$ is a program that takes a modulus $m \in \mathbb{N}^>$ and $n$ integers $a_0, \ldots, a_{n-1} \in \{0, \ldots, m-1\}$ as input, and that returns $f(a_0, \ldots, a_{n-1}) \operatorname{rem} m \in \mathbb{N}_m$. A *modular blackbox polynomial* is a polynomial $f$ that is represented in this way. The *cost* (or, better, a cost function) of such a polynomial is a function $\mathsf{C}_f$ such that $\mathsf{C}_f(s)$ yields an upper bound for the running time if $m$ has bit-size $\leqslant s$. It will be convenient to always assume that the *average cost* $\mathsf{A}_f(s) := \mathsf{C}_f(s)/s$ per bit of the modulus is non-decreasing and that $\mathsf{A}_f(ks) \leqslant k^{O(1)} \mathsf{A}_f(s)$ for any $k \geqslant 1$. Since $f$ should at least read its $n$ input values, we also assume that $n = O(\mathsf{A}_f(s))$.

**Remark 2.2.** A popular type of modular blackboxes are straight-line programs (SLPs) [10]. For an SLP of length $L$ that only uses ring operations, the above average cost function usually becomes $\mathsf{A}_f(s) \leqslant CL \log s$, for some fixed constant $C$ that does not depend on $f$.

If the SLP also allows for divisions, then we rather obtain $\mathsf{A}_f(s) \leqslant CL \log^2 s$, but this is out of the scope of this paper, due to the "division by zero" issue. In fact, computation trees [10] are more suitable than SLPs in this context. For instance, the computation of determinants using Gaussian elimination naturally fits in this setting, since the chosen computation path may then depend on the modulus $m$.

However, although these bounds "usually" hold (i.e. for all common algebraic algorithms that we are aware of, including the RAM model), they may fail in pathological cases when the SLP randomly accesses data that are stored at very distant locations on the Turing tape. For this reason, the blackbox cost model may be preferred in order to study bit complexities. In this model, a suitable replacement for the length $L$ of an SLP is the average cost function $A_f(s)$, which typically involves only a logarithmic overhead in the bit-length $s$.

## 2.3. Number theoretic reminders

All along this paper $T$ will bound the number of terms of the polynomial to be interpolated, and $r$ and $q$ will denote random prime numbers that satisfy:

- $T \leqslant r = O(T)$,
- $q \in \mathbb{N} r + 1$,
- $q = O(r^6)$.

The construction of $r$ and $q$ will rely on the following number theoretic theorems, where log stands for the natural logarithm, that is $\log e = 1$. We will also use $\log_2 x := \log x / \log 2$.

THEOREM 2.3. [42, Equation (3.8)] *For $\lambda \geqslant 21$, there exist at least $\frac{3}{5} \lambda / \log \lambda$ distinct prime numbers in the open interval $(\lambda, 2\lambda)$.*

THEOREM 2.4. *Let $\rho(x) := \frac{1.538 \log x}{\log \log x}$ if $x \geqslant e^e$ and $\rho(x) := 1.538 \, e$ otherwise. For all $N \geqslant 1$, the number of prime divisors of $N$ is bounded by $\rho(N)$.*

**Proof.** The function $\rho(x)$ is increasing for $x \geqslant e^e$ and $\rho(e^e) = 1.538 \, e$. So it is always non-decreasing and continuous. The number of prime divisor of any $N \leqslant 15$ it at most $2 \leqslant 1.538 \, e$. Let $d(N)$ and $\omega(N)$ respectively be the number of divisors and prime divisors of $N$. Then clearly $2^{\omega(N)} \leqslant d(N)$. Now for all $N \geqslant 3$ we know from [37] that

$$\omega(N) \leqslant \log_2 d(N) \leqslant \rho(N). \qquad \Box$$

We will need the following slightly modified version of [16, Theorem 2.1], which is itself based on a result from [43].

THEOREM 2.5. *There exists a Monte Carlo algorithm which, given $\varepsilon > 0$ and $R \geqslant \frac{2^{58}}{\varepsilon^2}$, produces a triple $(r, q, \omega)$ that has the following properties with probability at least $1 - \varepsilon$, and returns fail otherwise:*

a) *$r$ is uniformly distributed amongst the primes of $(R, 2R)$;*

b) *there are at least $R^5/(24 \log R)$ primes in $(2R, R^6) \cap (r \mathbb{N} + 1)$ and $q$ is uniformly distributed amongst them;*

c) *$\omega$ is a primitive $r$-th root of unity in $\mathbb{F}_q$.*

*Its worst-case bit complexity is $(\log R)^{O(1)}$.*

**Proof.** In [16, Theorem 2.1] the statement $(b)$ is replaced by the simpler condition that $q \leqslant R^6$. But by looking at step 2 of the algorithm on page 4 of [43], we observe that $q$ is actually uniformly distributed amongst the primes of $(2R, R^6) \cap (r\mathbb{N} + 1)$ and that there are at least $R^5/(24 \log R)$ such primes with high probability $\geqslant 1 - \frac{\varepsilon}{4}$. $\square$

LEMMA 2.6. *Let $\varepsilon$ be a real number in $(0,1)$ and let $P \geqslant 22$ be such that $P/\log P > 4n$. There exists a Monte Carlo algorithm that computes distinct random prime numbers $p_0, \ldots, p_{n-1}$ in $(P, 2P)$ in time*

$$O(n(\log n + \log(\varepsilon^{-1})))(\log P)^{O(1)},$$

*with a probability of success of at least $1 - \varepsilon$.*

**Proof.** Theorem 2.3 asserts that there are at least $\frac{3}{5} P / \log P$ primes in the interval $(P, 2P)$. The probability to fetch a prime number in $(P, 2P)$ while avoiding at most $n$ fixed numbers is at least

$$\frac{\frac{3}{5} \frac{P}{\log P} - n}{P} \geqslant \frac{\left(\frac{3}{5} - \frac{1}{4}\right) \frac{P}{\log P}}{P} = \frac{7}{20 \log P}.$$

The probability of failure after $k$ trials is at most $\left(1 - \frac{7}{20 \log P}\right)^k$. By using the AKS algorithm [1] each primality test takes time $(\log P)^{O(1)}$. The probability of success for picking $n$ distinct prime numbers in this way is at least

$$\left(1 - \left(1 - \frac{7}{20 \log P}\right)^k\right)^n.$$

In order to guarantee this probability of success to be at least $1 - \varepsilon$, it suffices to take

$$k \geqslant \frac{-\log(1 - (1-\varepsilon)^{1/n})}{-\log\left(1 - \frac{7}{20 \log P}\right)}.$$

The concavity of the log function yields $x \leqslant -\log(1-x) \leqslant \frac{x}{1-x}$ for $x \in (0,1)$, whence

$$\frac{-\log(1-\varepsilon)}{n} \geqslant \frac{\varepsilon}{n} \geqslant \frac{\frac{\varepsilon}{2n}}{1 - \frac{\varepsilon}{2n}} \geqslant -\log\left(1 - \frac{\varepsilon}{2n}\right),$$

and consequently,

$$-\log(1 - (1-\varepsilon)^{1/n}) \leqslant -\log\left(\frac{\varepsilon}{2n}\right).$$

On the other hand we have $-\log\left(1 - \frac{7}{20 \log P}\right) \geqslant \frac{7}{20 \log P}$. It therefore suffices to take

$$k := \left\lceil \frac{20}{7} (\log(2n) + \log(\varepsilon^{-1})) \log P \right\rceil. \quad \square$$

### 2.4. Amortized determination of prime divisors in a fixed set

Let $p_0 < \cdots < p_{n-1}$ be prime numbers and let $a_0 < \cdots < a_{N-1}$ be strictly positive integers. The aim of this subsection is to show that the set of pairs $\{(i,k) : i \in \mathbb{N}_n, k \in \mathbb{N}_N, p_i | a_k\}$ can be computed in quasi-linear time using the following algorithm named divisors.

**Algorithm divisors**
**Input:** non empty subsets $\mathcal{I} \subseteq \mathbb{N}_n$ and $\mathcal{K} \subseteq \mathbb{N}_N$.
**Output:** the set $\{(i,k) \in \mathcal{I} \times \mathcal{K} : p_i | a_k\}$.

1. If $|\mathcal{K}| = 1$, then return $\{(i,k) \in \mathcal{I} \times \mathcal{K} : p_i | a_k\}$.
2. Let $h := \lfloor |\mathcal{K}|/2 \rfloor$, let $\mathcal{K}_1$ be a subset of $\mathcal{K}$ of cardinality $h$, and let $\mathcal{K}_2 := \mathcal{K} \setminus \mathcal{K}_1$.
3. Compute $A_1 := \prod_{k \in \mathcal{K}_1} a_k$ and $A_2 := \prod_{k \in \mathcal{K}_2} a_k$.
4. Compute $\mathcal{I}_1 := \{i \in \mathcal{I} : p_i | A_1\}$ and $\mathcal{I}_2 := \{i \in \mathcal{I} : p_i | A_2\}$.
5. Return divisors($\mathcal{I}_1, \mathcal{K}_1$) $\cup$ divisors($\mathcal{I}_2, \mathcal{K}_2$).

LEMMA 2.7. *The algorithm* divisors *is correct and runs in time* $O(s \log^3 s)$, *where* $s := \log(p_0 \cdots p_{n-1} a_0 \cdots a_{N-1})$.

**Proof.** Let $\alpha := \log(\prod_{k \in \mathcal{K}} a_k)$ and $\beta := \log(\prod_{i \in \mathcal{I}} p_i)$. Step 1 costs $O((\alpha + \beta) \log^2(\alpha + \beta))$ by using fast multi-remaindering [15, Chapter 10]. Using fast sub-product trees [15, Chapter 10], step 3 takes $O(\alpha \log^2 \alpha)$. By means of fast multi-remaindering again, step 4 takes $O((\alpha + \beta) \log^2(\alpha + \beta))$.

Since the $p_i$ are distinct prime numbers, when entering step 5 we have

$$\prod_{i \in \mathcal{I}_m} p_i \leqslant \prod_{k \in \mathcal{K}_m} a_k \quad \text{for} \quad m = 1, 2.$$

Let $\mathsf{T}(\alpha)$ denote the cost of the recursive calls occurring during the execution of the algorithm. So far we have shown that

$$\mathsf{T}(\alpha) = \mathsf{T}(\log A_1) + \mathsf{T}(\log A_2) + O(\alpha \log^2 \alpha).$$

Unrolling this inequality and taking into account that the depth of the recursion is $O(\log N) = O(\log(a_0 \cdots a_{N-1}))$, we deduce that $\mathsf{T}(\alpha) = O(\alpha \log^3 \alpha)$. Finally the total cost of the algorithm is obtained by adding the cost of the top level call, that is

$$\mathsf{T}(\alpha) + O((\alpha + \beta) \log^2(\alpha + \beta)) = O(s \log^3 s). \qquad \square$$

## 3. PROBABILISTIC CODES FOR EXPONENTS

Consider a sparse polynomial $f = \sum_{e \in \mathbb{N}^n} c_e x^e$ that we wish to interpolate. In the next section, we will describe a method that allows us to efficiently compute most of the exponents $e$ in an encoded form $\phi(e)$. The simplest codes $\phi(e)$ are of the form

$$\phi(e) = e_0 \mu_0 + \cdots + e_{n-1} \mu_{n-1}. \tag{3.1}$$

When supp $f \subseteq \mathbb{N}_E^n$, the most common such encoding is the *Kronecker encoding*, with $\mu_i = E^i$ for all $i \in \mathbb{N}_n$. However, this encoding may incur large bit-size $\asymp n \sigma_E$ with respect to the bit-size of $e$.

The aim of this section is to introduce more compact codes $\phi(e)$. These codes will be "probabilistic" in the sense that we will only be able to recover $e$ from $\phi(e)$ with high probability, under suitable assumptions on $e$. Moreover, the recovery algorithm is only efficient if we wish to simultaneously "bulk recover" $T$ exponents from their codes.

## 3.1. The exponent encoding

Throughout this section, the number of variables $n \in \mathbb{N}^>$ and the target number of exponents $T \in \mathbb{N}^>$ are assumed to be given. We allow exponents to be vectors of arbitrary integers in $\mathbb{Z}^n$. Actual computations on exponents will be done modulo $B^\nu$ for a fixed odd base $B$ and a flexible $B$-adic precision $\nu$. We also fix a constant $P \geqslant 22$ such that

$$2n \log B < P \leqslant (2P)^2 < B \tag{3.2}$$

and we note that this assumption implies

$$\frac{P}{\log P} > 4n \tag{3.3}$$

and

$$B \geqslant 1937 \quad \text{and} \quad \sigma_B \geqslant 11. \tag{3.4}$$

We finally assume $\gamma \geqslant 1$ to be a parameter that will be specified in section 4. The exponent encoding will depend on one more parameter

$$1 \leqslant m \leqslant n$$

that will be fixed in section 3.2 and flexible in section 3.3. We define

$$\lambda := \left\lceil \gamma \frac{n}{m} \right\rceil.$$

Our encoding also depends on the following random parameters:

- For each $k \in \mathbb{N}_\lambda$, let $i_{k,0}, \ldots, i_{k,m-1}$ be random elements in $\mathbb{N}_n$ and $I_k := \{i_{k,0}, \ldots, i_{k,m-1}\}$.
- Let $p_0, \ldots, p_{n-1}$ be pairwise distinct random prime numbers in the interval $(P, 2P)$; such primes do exist thanks to Lemma 2.6 and (3.3).

Now consider an exponent $e = (e_0, \ldots, e_{n-1}) \in \mathbb{Z}^n$. We encode $e$ as

$$\phi_k(e) := \left(\sum_{i \in I_k} p_i e_i\right) \text{rem } B^\nu \in \mathbb{Z}_{B^\nu} \quad \text{for} \quad k \in \mathbb{N}_\lambda$$

$$\phi(e) := (\phi_0(e), \ldots, \phi_{\lambda-1}(e)) \in \mathbb{Z}_{B^\nu}^\lambda.$$

We will sometimes write $\phi^{[\nu,p,I]}$ and $\phi_k^{[\nu,p,I]}$ instead of $\phi$ and $\phi_k$ in order to make the dependence on $\nu$, $p$ and $I$ explicit.

## 3.2. Guessing individual exponents of prescribed size

Given an exponent $e = (e_0, \ldots, e_{n-1}) \in \mathbb{Z}^n$ of $f$, our first aim is to determine the individual exponents $e_i$ of small size. More precisely, assuming that

$$\#e := |\{i \in \mathbb{N}_n : e_i \neq 0\}| \leqslant \frac{n}{m},$$

we wish to determine all $e_i$ with $4P|e_i| < B^\nu$.

We say that $\phi(e)$ is *transparent* if for each $i \in \mathbb{N}_n$ with $e_i \neq 0$, there exists a $k \in \mathbb{N}_\lambda$ such that $\{j \in I_k : e_j \neq 0\} = \{i\}$. This property does only depend on the random choices of the $I_k$.

LEMMA 3.1. *Assume that $\#e \leqslant n/m$. Then, for random choices of $I_1, \ldots, I_\lambda$, the probability that $\phi(e)$ is transparent is at least $1 - (n/m) e^{-\gamma/e}$.*

**Proof.** Let $\mathcal{I} := \{i \in \mathbb{N}_n : e_i \neq 0\}$ and $\#e = |\mathcal{I}| \leq n/m$. Given $k \in \mathbb{N}_\lambda$ and $i \in \mathcal{I}$, the probability that $I_k \cap \mathcal{I} = \{i\}$ is

$$\frac{m(n-\#e)^{m-1}}{n^m} = \frac{m}{n}\left(1-\frac{\#e}{n}\right)^{m-1} \geq \frac{m}{n}\left(1-\frac{1}{m}\right)^{m-1} \geq e^{-1}\frac{m}{n}.$$

The probability that $I_k \cap \mathcal{I} = \{i\}$ for some $k$ is therefore at least

$$1 - \left(1 - e^{-1}\frac{m}{n}\right)^{\gamma\frac{n}{m}} \geq 1 - e^{-\gamma/e}.$$

We conclude that the probability that this holds for all $i \in \mathcal{I}$ is at least $1 - (n/m)e^{-\gamma/e}$. $\square$

We say that $\phi(e)$ is *faithful* if for every $k \in \mathbb{N}_\lambda$ and $i \in \mathbb{N}_n$ such that $4P|e_i| < B^\nu$ and $p_i | \phi_k(e)$, we have $\phi_k(e) = p_i e_i$.

LEMMA 3.2. *For random choices of $p_0, \ldots, p_{n-1}$, the code $\phi(e)$ is faithful with probability at least*

$$1 - \frac{16\lambda\nu n^2 \log B}{P}.$$

**Proof.** Let $\mathcal{P}$ be the set of all primes strictly between $P$ and $2P$. Let $\mathcal{U}$ be the set of all $(p_0, \ldots, p_{n-1}) \in \mathcal{P}^n$ such that $p_0, \ldots, p_{n-1}$ are pairwise distinct. Let $\mathcal{X}$ be the subset of $\mathcal{U}$ of all choices of $p_0, \ldots, p_{n-1}$ for which $\phi(e)$ is not faithful.

Consider $k \in \mathbb{N}_\lambda$, $i \in \mathbb{N}_n$, and $(p_0, \ldots, p_{n-1}) \in \mathcal{U}$ be such that $p_i | \phi_k(e)$, $4P|e_i| < B^\nu$ and $\phi_k(e) \neq p_i e_i$. Let $\Phi := \phi_k(e) - p_i e_i \neq 0$. For each $q \in \mathcal{P}$, let

$$p_{i \to q} := (p_0, \ldots, p_{i-1}, q, p_{i+1}, \ldots, p_{n-1})$$

and $\phi^{[q]} := \phi^{[\nu, p_{i \to q}, I]}$, so that $\phi^{[p_i]} = \phi$. For each $q \in \mathcal{P}$, using $4P|e_i| < B^\nu$, we observe that

$$\phi_k^{[q]}(e) = \Phi + q e_i + \epsilon B^\nu$$

necessarily holds with $\epsilon \in \{-1, 0, 1\}$.

Now consider the set $Q_{i,k,p_1,\ldots,p_{i-1},p_{i+1},\ldots,p_n}$ of $q \in \mathcal{P}$ such that $\phi_k^{[q]}(e)$ is divisible by $q$. Any such $q$ is a divisor of either $\Phi - B^\nu$, $\Phi$, or $\Phi + B^\nu$. Since $B$ is odd we have

$$|\phi_k(e)| \leq \frac{B^\nu - 1}{2},$$

and therefore

$$|\Phi| \leq |\phi_k(e)| + |p_i e_i| \leq \frac{B^\nu - 1}{2} + \frac{B^\nu}{2} = B^\nu - \frac{1}{2}.$$

It follows that $\Phi + \varepsilon B^\nu \neq 0$. Consequently,

$$\begin{aligned}
|Q_{i,k,p_1,\ldots,p_{i-1},p_{i+1},\ldots,p_n}| &\leq 3\left\lceil\frac{\log(2B^\nu)}{\log P}\right\rceil \\
&\leq 3\left(\frac{\log(2B^\nu)}{\log P} + 1\right) \\
&= 3\left(\frac{\nu \log B}{\log P} + \frac{\log(2P)}{\log P}\right) \\
&\leq 3\left(\frac{\nu \log B}{\log P} + \frac{\log B}{2\log P}\right) \qquad \text{(by (3.2))} \\
&\leq \frac{9\nu \log B}{2\log P}.
\end{aligned}$$

Now let
$$\mathcal{X}_{i,k} := \{(p_0, \ldots, p_{n-1}) \in \mathcal{U} : p_i \in \mathcal{Q}_{i,k,p_0,\ldots,p_{i-1},p_{i+1},\ldots,p_{n-1}}\},$$
so that $\mathcal{X} \subseteq \bigcup_{i,k} \mathcal{X}_{i,k}$. By what precedes, we have
$$|\mathcal{X}_{i,k}| \leqslant \frac{9}{2} \binom{|\mathcal{P}|}{n-1} \frac{\nu \log B}{\log P},$$
whence
$$|\mathcal{X}| \leqslant \frac{9}{2} \lambda \nu n \binom{|\mathcal{P}|}{n-1} \frac{\log B}{\log P}.$$
From $|\mathcal{U}| = \binom{|\mathcal{P}|}{n}$ we deduce that
$$\frac{|\mathcal{X}|}{|\mathcal{U}|} \leqslant \frac{9 \lambda \nu n^2 \log B}{2(|\mathcal{P}| - n + 1) \log P}.$$
From Theorem 2.3 we know that $|\mathcal{P}| \geqslant 3/5 \, P / \log P$. This yields $|\mathcal{P}| \geqslant 9/16 \, P / \log P$, as well as $|\mathcal{P}| \geqslant 2n$, thanks to (3.3). We conclude that
$$\frac{|\mathcal{X}|}{|\mathcal{U}|} \leqslant \frac{9 \lambda \nu n^2 \log B}{|\mathcal{P}| \log P} \leqslant \frac{16 \lambda \nu n^2 \log B}{P}. \qquad \square$$

Given $\psi \in \mathbb{Z}_{B^\nu}^\lambda$ and $i \in \mathbb{N}_n$, let $k_{\psi,i}$ be the smallest index such that $p_i \mid \psi_{k_{\psi,i}} \neq 0$ and $4P |\psi_{k_{\psi,i}}|/p_i < B^\nu$. If no such $k_{\psi,i}$ exists, then we let $k_{\psi,i} := \bot$. We define $k_\psi := (k_{\psi,0}, \ldots, k_{\psi,n-1})$.

Assume that $\psi = \phi(e)$ is transparent and faithful for some $e \in \mathbb{Z}^n$. Given $i \in \mathbb{N}_n$, let $\tilde{e}_i := \psi_{k_{\psi,i}}/p_i$ if $k_{\psi,i} \neq \bot$ and $\tilde{e}_i := 0$ otherwise. Then the condition $4P|\psi_{k_{\psi,i}}|/p_i < B^\nu$ implies that $2p_i|\tilde{e}_i| < B^\nu$ always holds. Moreover, if $4P|e_i| < B^\nu$, then the definitions of "transparent" and "faithful" imply that $e_i = \tilde{e}_i$. In other words, these $e_i$ can efficiently be recovered from $\psi$ and $k_\psi$.

LEMMA 3.3. *Let $(\psi^0, \ldots, \psi^{T-1}) \in (\mathbb{Z}_{B^\nu}^\lambda)^T$, where $T\nu \geqslant m$. Then we can compute $(k_{\psi^0}, \ldots, k_{\psi^{T-1}})$ in time $\tilde{O}(T \lambda \nu \log B)$.*

**Proof.** Note that the hypotheses $T\nu \geqslant m$ and $\gamma \geqslant 1$ imply that $n = O(T \lambda \nu)$. Using Lemma 2.7, we can compute all triples $(j, i, k)$ with $p_i \mid \psi_k^j$ in time
$$\tilde{O}(T \lambda \nu \log B + n \log P) = \tilde{O}(T \lambda \nu \log B),$$
thanks to (3.2). Using $\tilde{O}(T \lambda \nu \log B)$ further operations we can filter out the triples $(j, i, k)$ for which $4P |\psi_k^j|/p_i < B^\nu$. We next sort the resulting triples for the lexicographical ordering, which can again be done in time $\tilde{O}(T \lambda \nu \log B)$. For each pair $(j, i)$, we finally only retain the first triple of the form $(j, i, k)$. This can once more be done in time $\tilde{O}(T \lambda \nu \log B)$. Then the remaining triples are precisely those $(j, i, k_{\psi^j,i})$ with $k_{\psi^j,i} \neq \bot$. $\square$

The following combination of the preceding lemmas will be used in the sequel:

LEMMA 3.4. *Let $e = (e^0, \ldots, e^{T-1}) \in (\mathbb{Z}^n)^T$ and let $\boldsymbol{\psi} = (\psi^0, \ldots, \psi^{T-1}) \in (\mathbb{Z}_{B^\nu}^\lambda)^T$ be the corresponding values as above. Let $J$ be the set of indices $j \in \mathbb{N}_T$ such that $\psi^j = \phi(e^j)$ and $\#e^j \leqslant n/m$. Then there exists an algorithm that takes $\boldsymbol{\psi}$ as input and that computes $\tilde{e} = (\tilde{e}^0, \ldots, \tilde{e}^{T-1}) \in (\mathbb{Z}^n)^T$ such that $2p_i|\tilde{e}_i^j| < B^\nu$ for all $(i,j) \in \mathbb{N}_n \times \mathbb{N}_T$ and $\tilde{e}_i^j = e_i^j$ for all $j \in J$ with $4P|e_i^j| < B^\nu$. The algorithm runs in time $\tilde{O}(T \lambda \nu \log B)$ and succeeds with probability at least*
$$1 - T \frac{n}{m} e^{-\gamma/e} - \frac{16 \lambda \nu T n^2 \log B}{P}.$$

**Proof.** Lemmas 3.1 and 3.2 bound the probability of failure for the transparency and the faithfulness of the random choices for each of the $T$ terms. The complexity bound follows from Lemma 3.3. □

### 3.3. Guessing exponents of prescribed size

Our next aim is to determine all exponents $e^j$ with $\sigma_{e^j} \leqslant \Sigma$, for some fixed threshold $\Sigma$. For this, we will apply Lemma 3.4 several times, for different choices of $\phi^{[\nu,p,I]}$. Let

$$U := \lceil \log_2 \min(\Sigma, n) \rceil + 2.$$

For $u = 1, \ldots, U$, let

$$\nu^{\{u\}} := \left\lceil \frac{5\Sigma}{2^{U-u}} \right\rceil$$

$$m^{\{u\}} := \left\lceil \frac{n}{2^{U-u}} \right\rceil$$

$$\lambda^{\{u\}} := \left\lceil \gamma \frac{n}{m^{\{u\}}} \right\rceil.$$

We also choose $p^{\{u\}}$ and $I^{\{u\}}$ independently at random as explained in section 3.1, with $\nu^{\{u\}}$, $m^{\{u\}}$, and $\lambda^{\{u\}}$ in the roles of $\nu$, $m$, and $\lambda$. We finally define

$$\phi^{\{u\}} := \phi^{[\nu^{\{u\}}, p^{\{u\}}, I^{\{u\}}]},$$

for $u = 1, \ldots, U$.

Note that the above definitions imply

$$4 \min(\Sigma, n) \leqslant 2^U < 8 \min(\Sigma, n) \qquad (3.5)$$

and $1 \leqslant m^{\{u\}} \leqslant n$. The inequality $m^{\{u\}} < n/2^{U-u} + 1$ implies

$$2^{U-u} \leqslant \frac{2n}{m^{\{u\}}}, \quad \text{whenever } m^{\{u\}} \geqslant 2. \qquad (3.6)$$

If $m^{\{u\}} = 1$, then $2^{U-u} \leqslant 2^{U-1} \leqslant 4n = 4n/m^{\{u\}}$, so, in general,

$$2^{U-u} \leqslant \frac{4n}{m^{\{u\}}}. \qquad (3.7)$$

By (3.5) we have $5\Sigma \geqslant 2^{U-1}$, whence

$$\nu^{\{1\}} < \nu^{\{2\}} < \cdots < \nu^{\{U\}}. \qquad (3.8)$$

LEMMA 3.5. *For $u = 1, \ldots, U$, we have $\lambda^{\{u\}} \nu^{\{u\}} \leqslant 18 \gamma \Sigma$.*

**Proof.** From $\gamma \geqslant 1$, we get

$$\lambda^{\{u\}} \leqslant \lceil \gamma 2^{U-u} \rceil \leqslant \gamma 2^{U-u+1}. \qquad (3.9)$$

Now

$$\lambda^{\{u\}} \nu^{\{u\}} < \left( \frac{5\Sigma}{2^{U-u}} + 1 \right) \gamma 2^{U-u+1} \qquad \text{(by (3.9))}$$

$$\leqslant \gamma (10\Sigma + 2^{U-u+1})$$

$$\leqslant 18 \gamma \Sigma, \qquad \text{(by (3.5))}$$

which concludes the proof. □

THEOREM 3.6. *Let $e = (e^0, \ldots, e^{T-1}) \in (\mathbb{Z}^n)^T$ and let $\psi^{\{u\}} = (\psi^{\{u,0\}}, \ldots, \psi^{\{u,T-1\}}) \in (\mathbb{Z}_{B^{\nu\{u\}}}^\lambda)^T$ for $i = 1, \ldots, U$ be as above. Assume that $T\Sigma \geqslant n$. Let $J$ be the set of indices $j \in \mathbb{N}_T$ such that $\sigma_{e^j} \leqslant \Sigma$ and $\psi^{\{u,j\}} = \phi^{\{u\}}(e^j)$ for all $u = 1, \ldots, U$. Then there exists an algorithm that takes $\psi^{\{1\}}, \ldots, \psi^{\{u\}}$ as input and that computes $\hat{e} = (\hat{e}^0, \ldots, \hat{e}^{T-1}) \in (\mathbb{Z}^n)^T$ such that $\sigma_{\hat{e}^j} \leqslant \Sigma$ for all $j \in \mathbb{N}_T$ and $\hat{e}^j = e^j$ for all $j \in J$. The algorithm runs in time $\tilde{O}(\gamma T \Sigma \log B)$ and succeeds with probability at least*

$$1 - TnU\mathrm{e}^{-\gamma/\mathrm{e}} - \frac{288\gamma U\Sigma Tn^2 \log B}{P}.$$

**Proof.** We compute successive approximations $e^{\{0\}}, e^{\{1\}}, \ldots, e^{\{U\}}$ of $e$ as follows. We start with $e^{\{0\}} := 0$. Assuming that $e^{\{u-1\}}$ is known, we compute

$$\dot{\psi}^j := (\psi^{\{u,j\}} - \phi^{\{u\}}((e^{\{u-1\}})^j)) \operatorname{rem} B^{\nu^{\{u\}}} \in \mathbb{Z}_{B^{\nu^{\{u\}}}}^\lambda$$

for all $j \in \mathbb{N}_T$, and then apply the algorithm underlying Lemma 3.4 to $\dot{e} := e - e^{\{u-1\}}$ and $\dot{\psi} := (\dot{\psi}^0, \ldots, \dot{\psi}^{T-1})$. Note that for all $j \in J$ the equality $\psi^{\{u,j\}} = \phi^{\{u\}}(e^j)$ implies

$$\begin{aligned}
\dot{\psi}^j &= (\phi^{\{u\}}(e^j) - \phi^{\{u\}}((e^{\{u-1\}})^j)) \operatorname{rem} B^{\nu^{\{u\}}} \\
&= \phi^{\{u\}}(e^j - (e^{\{u-1\}})^j) \\
&= \phi^{\{u\}}(\dot{e}^j).
\end{aligned}$$

Our choice of $\nu^{\{u\}}$ and $m^{\{u\}}$, the inequality (3.5), and the assumption $T\Sigma \geqslant n$ successively ensure that

$$\frac{m^{\{u\}}}{\nu^{\{u\}}} \leqslant \frac{\frac{n}{2^{U-u}}+1}{\frac{5\Sigma}{2^{U-u}}} \leqslant \frac{n+2^{U-1}}{5\Sigma} \leqslant \frac{n}{\Sigma} \leqslant T,$$

so we can actually apply Lemma 3.4.

Let $J^{\{u\}}$ be the set of indices $j \in \mathbb{N}_T$ such that $\dot{\psi}^j = \phi^{\{u\}}(\dot{e}^j)$ and $\#\dot{e}^j \leqslant n/m^{\{u\}}$. Lemma 3.4 provides us with $\tilde{e} = (\tilde{e}^0, \ldots, \tilde{e}^{T-1}) \in (\mathbb{Z}^n)^T$ such that $2p_i |\tilde{e}_i^j| < B^{\nu^{\{u\}}}$ for all $(i,j) \in \mathbb{N}_n \times \mathbb{N}_T$. In addition $\tilde{e}_i^j = \dot{e}_i^j$ holds whenever $j \in J^{\{u\}}$ and $4P|\dot{e}_i^j| < B^{\nu^{\{u\}}}$ with probability at least

$$1 - T\frac{n}{m^{\{u\}}}\mathrm{e}^{-\gamma/\mathrm{e}} - \frac{16\lambda\nu^{\{u\}}Tn^2\log B}{P}.$$

Now we set

$$e^{\{u\}} := e^{\{u-1\}} + \tilde{e}. \tag{3.10}$$

At the end, we return $\hat{e} := e^{\{U\}}$ as our guess for $e$. For the analysis of this algorithm, we first assume that all applications of Lemma 3.4 succeed. Let us show by induction (and with the above definitions of $\dot{\psi}$ and $\dot{e}$) that, for all $i \in \mathbb{N}_n$, $j \in J$, and $u = 1, \ldots, U$, we have:

  i. $\#\dot{e}^j \leqslant n/m^{\{u\}}$;

 ii. $2P|(e^{\{u\}})_i^j| < B^{\nu^{\{u\}}}$;

iii. $(e^{\{u\}})_i^j = e_i^j$ whenever $4P|e_i^j| < B^{\nu^{\{u\}}}\left(1 - \frac{2}{B}\right)$.

If $u = 1$ and $m^{\{1\}} = 1$ then (i) clearly holds. If $u = 1$ and $m^{\{1\}} \geqslant 2$, then (3.5) and (3.6) imply $n/m^{\{1\}} \geqslant 2^{U-2} \geqslant \min(\Sigma, n)$. Since $j \in J$ we have $\sigma_{e^j} \leqslant \Sigma$. Now we clearly have $\#\dot{e}^j \leqslant \sigma_{e^j}$ and $\#\dot{e}^j \leqslant n$, so (i) holds.

If $u \geqslant 2$, then let $i \in \mathbb{N}_n$ be such that $\dot{e}_i^j \neq 0$, so we have $(e^{\{u-1\}})_i^j \neq e_i^j$. The induction hypothesis (iii) and (3.4) yield

$$4P|e_i^j| \geqslant B^{\nu^{\{u-1\}}}\left(1-\frac{2}{B}\right) \geqslant \frac{2}{3}B^{\nu^{\{u-1\}}},$$

whence $6P|e_i^j| \geqslant B^{\nu^{\{u-1\}}}$. Consequently,

$$\begin{aligned}
\log_2|e_i^j| &\geqslant \nu^{\{u-1\}}\log_2 B - \log_2(6P) \\
&\geqslant \nu^{\{u-1\}}\log_2\frac{B}{6P} &&\text{(since } \nu^{\{u-1\}} \geqslant 1\text{)} \\
&\geqslant \nu^{\{u-1\}}\log_2\frac{\sqrt{B}}{3} &&\text{(by (3.2))} \\
&\geqslant \frac{\nu^{\{u-1\}}(\sigma_B-3)}{2} &&\text{(since } B \geqslant 2^{\sigma_B-1}\text{)} \\
&\geqslant \frac{5\Sigma(\sigma_B-3)}{2^{U-u+2}} \\
&\geqslant \frac{40\Sigma}{2^{U-u+2}} &&\text{(by (3.4))} \\
&\geqslant \frac{\Sigma}{n/m^{\{u\}}}. &&\text{(by (3.7))}
\end{aligned}$$

Hence the total bit-size of all $|e_i^j|$ such that $\dot{e}_i^j \neq 0$ is at least $(\#\dot{e}^j)\Sigma/(n/m^{\{u\}})$ and at most $\Sigma$ by definition of $J$. This concludes the proof of (i). Now if $j \in J$, then $\psi^{\{u,j\}} = \phi^{\{u\}}(e^j) = \dot{\psi}^j$, so Lemma 3.4 and (i) imply that $j \in J^{\{u\}}$. In other words, we obtain an approximation $\tilde{e}$ of $\dot{e}$ such that $2p_i^{\{u\}}|\tilde{e}_i^j| < B^{\nu^{\{u\}}}$ for all $(i,j) \in \mathbb{N}_n \times J$, and $\tilde{e}_i^j = \dot{e}_i^j$ holds whenever $4P|\dot{e}_i^j| < B^{\nu^{\{u\}}}$.

Let us prove (ii). If $u=1$, then $2P|(e^{\{1\}})_i^j| = 2P|\tilde{e}_i^j| < B^{\nu^{\{u\}}}$. If $u \geqslant 2$, then (3.10) yields

$$\begin{aligned}
2P|(e^{\{u\}})_i^j| &\leqslant 2P|(e^{\{u-1\}})_i^j| + 2P|\tilde{e}_i^j| \\
&< B^{\nu^{\{u-1\}}} + 2P|\tilde{e}_i^j| &&\text{(by induction hypothesis (ii))} \\
&\leqslant B^{\nu^{\{u-1\}}} + \frac{2P}{2p_i^{\{u\}}}B^{\nu^{\{u\}}} \\
&\leqslant B^{\nu^{\{u-1\}}} + \frac{P}{P+1}B^{\nu^{\{u\}}} &&\text{(since } p_i^{\{u\}} \geqslant P+1\text{)} \\
&\leqslant \frac{B^{\nu^{\{u\}}}}{B} + \frac{P}{P+1}B^{\nu^{\{u\}}} &&\text{(by (3.8))} \\
&\leqslant \frac{B^{\nu^{\{u\}}}}{P+1} + \frac{P}{P+1}B^{\nu^{\{u\}}} &&\text{(by (3.2))} \\
&= B^{\nu^{\{u\}}}.
\end{aligned}$$

As to (iii), assume that $4P|e_i^j| < B^{\nu^{\{u\}}}\left(1-\frac{2}{B}\right)$. If $u=1$, then $(e^{\{u\}})_i^j = \dot{e}_i^j = e_i^j$ is immediate. If $u \geqslant 2$, then the induction hypothesis (ii) yields $2P|(e^{\{u-1\}})_i^j| < B^{\nu^{\{u-1\}}}$, whence

$$\begin{aligned}
4P|\dot{e}_i^j| &\leqslant 4P|e_i^j| + 4P|(e^{\{u-1\}})_i^j| \\
&< B^{\nu^{\{u\}}} - 2B^{\nu^{\{u\}}-1} + 2B^{\nu^{\{u-1\}}} \\
&\leqslant B^{\nu^{\{u\}}}. &&\text{(by (3.8))}
\end{aligned}$$

We deduce that $\tilde{e}_i^j = \dot{e}_i^j$ holds, hence (3.10) implies (iii). This completes our induction.

At the end of the induction, we have $\nu^{\{U\}} = 5\Sigma$ and, for all $(i,j) \in \mathbb{N}_n \times J$,

$$\begin{aligned}
5P|e_i^j| &\leq 5P 2^\Sigma & \text{(since } \sigma_{e^j} \leq \Sigma\text{)} \\
&\leq {}^5\!/_2 B^{1/2} 2^\Sigma & \text{(by (3.2))} \\
&= {}^5\!/_2 B^{\nu^{\{U\}}/(5\log_2 B)+1/2} \\
&\leq {}^5\!/_2 B^{\nu^{\{U\}}/50+1/2} & \text{(by (3.4))} \\
&\leq B^{\nu^{\{U\}}/50+3/4} & \text{(by (3.4))} \\
&\leq B^{\nu^{\{U\}}}. & \text{(since } \nu^{\{U\}} \geq 1\text{)}
\end{aligned}$$

By (iii) and (3.4), this implies the correctness of the overall algorithm.

As to the complexity bound, Lemma 3.5 shows that $\lambda^{\{u\}} \nu^{\{u\}} = O(\gamma \Sigma)$, when applying Lemma 3.4. Consequently, the cost of all these applications for $u = 1, \ldots, U$ is bounded by

$$\tilde{O}\left(\sum_{u=1}^U \lambda^{\{u\}} T \nu^{\{u\}} \log B\right) = \tilde{O}(\gamma T \Sigma U \log B) = \tilde{O}(\gamma T \Sigma \log B),$$

since $U = O(\log \Sigma)$. The cost of the evaluations of $\phi^{\{u\}}((e^{\{u-1\}})^j)$ and all further additions, subtractions, and modular reductions is not more expensive.

The bound for the probability of success follows by induction from Lemmas 3.4 and 3.5, while using the fact that all $p^{\{u\}}$ and $I^{\{u\}}$ are chosen independently. □

## 4. SPARSE INTERPOLATION

Throughout this section, we assume that $f$ is a modular blackbox polynomial in $\mathbb{Z}[x_1, \ldots, x_n]$ with at most $T$ terms and of bit-size at most $S \geq T$. In order to simplify the cost analyses, it will be convenient to assume that

$$S \geq \max(n, 2^{16}).$$

Our main goal is to interpolate $f$. From now on $\#f$ will stand for the actual number of non-zero terms in the sparse representation of $f$.

Our strategy is based on the computation of increasingly good approximations of the interpolation of $f$, as in [3], for instance. The idea is to determine an approximation $\tilde{f}$ of $f$, such that $f - \tilde{f}$ contains roughly half the number of terms as $f$, and then to recursively re-run the same algorithm for $f - \tilde{f}$ instead of $f$. Our first approximation will only contain terms of small bit-size. During later iterations, we will include terms of larger and larger bit-sizes.

Throughout this section, we set

$$\beta := 2^6 \sigma_S^2 \quad \text{and} \quad \Sigma := \lfloor \beta S/T \rfloor, \tag{4.1}$$

so that at most $T/\beta$ of the terms of $f$ have size $>\Sigma$. Our main technical aim will be to determine at least $\max(T/2 - \#f, 0)$ terms of $f$ of size $\leq \Sigma$, with high probability.

### 4.1. Cyclic modular projections

Our interpolation algorithm is based on an extension of the univariate approach from [17]. One first key ingredient is to homomorphically project the polynomial $f$ to an element of $\mathbb{Z}[t]/(t^r - 1, M)$ for a suitable cycle length $r \in \mathbb{N}^>$ and a suitable modulus $M$ (of the form $M = B^\nu$, where $B$ is as in the previous section and $\nu \in \mathbb{N}^>$).

More precisely, we fix

$$R := \max(S, 2^{58}) \beta^2 \qquad (4.2)$$

and compute $(r, q, \omega)$ as in Theorem 2.5. Now let $\tau_0, \ldots, \tau_{n-1}$ be random elements of $\{1, \ldots, r-1\}$, and consider the map

$$\begin{aligned}\Pi_{\tau,r} \colon \mathbb{Z}[x_0, \ldots, x_{n-1}] &\longrightarrow \mathbb{Z}[t]/(t^r - 1) \\ f &\longmapsto f(t^{\tau_0}, \ldots, t^{\tau_{n-1}}) \bmod (t^r - 1).\end{aligned}$$

We call $\Pi_{\tau,r}$ a *cyclic projection*.

LEMMA 4.1. *The bit-size of the product of the non-zero coefficients of $\Pi_{\tau,r}(f)$ is at most $\sigma_f$.*

**Proof.** Given $A = a_0 + \cdots + a_{r-1} t^{r-1} \in \mathbb{Z}[t]/(t^r - 1)$, let $\sigma_A := \sigma_{a_0} + \cdots + \sigma_{a_{r-1}}$ and $\pi(A) := \prod_{a_i \neq 0} a_i$. Note that $\sigma_{i+j} \leq \sigma_i + \sigma_j$ and $\sigma_{ij} \leq \sigma_i + \sigma_j$ for any $i, j \in \mathbb{Z}$. The first inequality yields $\sigma_{\Pi_{\tau,r}(f)} \leq \sigma_f$, whereas the second one implies $\sigma_{\pi(\Pi_{\tau,r}(f))} \leq \sigma_{\Pi_{\tau,r}(f)}$. □

Given a modulus $M \in \mathbb{N}^>$, we also define

$$\begin{aligned}\Pi_{\tau,r,M} \colon \mathbb{Z}[x_0, \ldots, x_{n-1}] &\longrightarrow \mathbb{Z}[t]/(t^r - 1, M) \\ f &\longmapsto f(t^{\tau_0}, \ldots, t^{\tau_{n-1}}) \bmod (t^r - 1, M)\end{aligned}$$

and call $\Pi_{\tau,r,M}$ a *cyclic modular projection*.

If $\Pi = \Pi_{\tau,r}$, then we say that a term $c x^e$ of $f$ and the corresponding exponent $e$ are $\Pi$-*faithful* if there is no other term $c' x^{e'}$ of $f$ such that $\Pi(x^e) = \Pi(x^{e'})$. If $\Pi = \Pi_{\tau,r,M}$, then we define $\Pi$-faithfulness in the same way, while requiring in addition that $c$ be invertible modulo $M$. For any $\kappa \geq 2$, we note that $c x^e$ is $\Pi_{\tau,r,M}$-faithful if and only if $c x^e$ is $\Pi_{\tau,r,M^\kappa}$-faithful. We say that $f$ is $\Pi$-faithful if all its terms are $\Pi$-faithful. In a similar way, we say that $\bar{f} := \Pi_{\tau,r}(f)$ is $M$-faithful if $c \operatorname{rem} M$ is invertible for any non-zero term $c t^{\bar{e}}$ of $\bar{f}$.

The first step of our interpolation algorithm is similar to the one from [17] and consists of determining the exponents of $\bar{f} := \Pi_{\tau,r}(f)$. Let $q$ be a prime number. If $\bar{f}$ is $q$-faithful, then the exponents of $\bar{f}$ are precisely those of $\bar{f} \operatorname{rem} q = \Pi_{\tau,r,q}(f)$.

LEMMA 4.2. *We can compute $\Pi_{\tau,r,q}(f)$ in time*

$$\mathsf{A}_f(\sigma_q) r \sigma_q + n \tilde{O}(r \log q).$$

**Proof.** We first precompute $1, \omega, \ldots, \omega^{r-1}$ in time $r \tilde{O}(\log q)$. We compute $\pi := \Pi_{\tau,r,q}(f)$ by evaluating $\pi(\omega^i) = f(\omega^{i \tau_0}, \ldots, \omega^{i \tau_{n-1}})$ for $i = 0, \ldots, r-1$. This takes $\mathsf{A}_f(\sigma_q) r \sigma_q + n r \tilde{O}(\log q)$ bit-operations. We next retrieve $\pi$ from these values using an inverse discrete Fourier transform (DFT) of order $r$. This takes $\tilde{O}(r)$ further operations in $\mathbb{F}_q$, using Bluestein's method [7]. □

Assuming that $\bar{f}$ is $q$-faithful and that we know $\Pi_{\tau,r,q}(f)$, consider the computation of $\Pi_{\tau,r,q^\nu}(f)$ for higher precisions $\nu$. Now $\bar{f}$ is also $q^\nu$-faithful, so the exponents of $\Pi_{\tau,r,q^\nu}(f)$ and $\Pi_{\tau,r,q}(f)$ coincide. One crucial idea from [17] is to compute $\Pi_{\tau,r,q^\nu}(f)$ using only $T' := \#\Pi_{\tau,r,q}(f)$ instead of $r$ evaluations of $f$ modulo $q^\nu$. This is the purpose of the next lemma.

**LEMMA 4.3.** *Assume that $\bar{f}$ is q-faithful and that $\Pi_{\tau,r,q}(f)$ is known. Let $T' := \#\Pi_{\tau,r,q}(f) \leqslant T$ and let $g(x) := f(\alpha x)$, where $\alpha = (\alpha_0, \ldots, \alpha_{n-1}) \in \mathbb{N}_{q^\nu}^n$ is such that $\alpha_i \operatorname{rem} q \neq 0$ for all $i \in \mathbb{N}_n$. Then we may compute $\Pi_{\tau,r,q^\nu}(g)$ in time*

$$\mathsf{A}_f(\nu \sigma_q) T' \nu \sigma_q + n \tilde{O}((T' + \log r) \nu \log q).$$

**Proof.** We first Hensel lift the primitive $r$-th root of unity $\omega$ in $\mathbb{F}_q$ to a principal $r$-th root of unity $\tilde{\omega}$ in $\mathbb{Z}/q^\nu \mathbb{Z}$ in time $\tilde{O}(\nu \log q \log r)$, as detailed in [17, section 2.2]. We next compute $\tilde{\omega}^{\tau_0}, \ldots, \tilde{\omega}^{\tau_{n-1}}$ in time $n\tilde{O}(\nu \log q \log r)$, using binary powering. We pursue with the evaluations $v_i := f(\alpha_0 \tilde{\omega}^{i\tau_0}, \ldots, \alpha_{n-1} \tilde{\omega}^{i\tau_{n-1}})$ for $i = 0, \ldots, T'-1$. This can be done in time

$$\mathsf{A}_f(\nu \sigma_q) T' \nu \sigma_q + nT' \tilde{O}(\nu \log q).$$

Now the exponents $e_0, \ldots, e_{T'} \in \mathbb{N}_r$ of $\Pi_{\tau,r,q^\nu}(g)$ are known, since they coincide with those of $\Pi_{\tau,r,q}(f)$, and we have the relation

$$\begin{pmatrix} 1 & 1 & \cdots & 1 \\ \tilde{\omega}^{e_0} & \tilde{\omega}^{e_1} & \cdots & \tilde{\omega}^{e_{T'-1}} \\ \vdots & \vdots & & \vdots \\ \tilde{\omega}^{(T'-1)e_0} & \tilde{\omega}^{(T'-1)e_1} & \cdots & \tilde{\omega}^{(T'-1)e_{T'-1}} \end{pmatrix} \begin{pmatrix} c_0 \\ c_1 \\ \vdots \\ c_{T'-1} \end{pmatrix} = \begin{pmatrix} v_0 \\ v_1 \\ \vdots \\ v_{T'-1} \end{pmatrix},$$

where $c_i$ denotes the coefficient of $t^{e_i}$ in $\Pi_{\tau,r,q^\nu}(g)$, for $i = 0, \ldots, T'-1$. It is well known that this linear system can be solved in quasi-linear time $\tilde{O}(T' \nu \log q)$: in fact this problem reduces to the usual interpolation problem thanks to the transposition principle [8, 33]; see for instance [24, section 5.1]. □

## 4.2. Probability of faithfulness

The next lemma is devoted to bounding the probability of picking up random prime moduli that yield $q$-faithful projections.

**LEMMA 4.4.** *Let $N$ (resp. $N^*$) be the number of terms (resp. $\Pi_{\tau,r,q}$-faithful terms) of $f$ of size $\leqslant \Sigma$. Let the cycle length $r$ and the modulus $q$ be given as described in Theorem 2.5. If the $\tau_0, \ldots, \tau_{n-1}$ are uniformly and independently taken at random $\{1, \ldots, r-1\}$, then, with probability $\geqslant 1 - \frac{3}{\sqrt{\beta}}$, the projection $\bar{f}$ is q-faithful and*

$$N^* \geqslant \left(1 - \frac{3}{\sqrt{\beta}}\right) N - \frac{T}{\beta}.$$

**Proof.** We let $\varepsilon := 1/\beta$ and $R \geqslant 2^{58}/\varepsilon^2$ as in (4.2), which allows us to apply Theorem 2.5. Let $f = \sum_e c_e x^e$ and $E := \{e \in \mathbb{N}^n : c_e \neq 0 \wedge \sigma_{c_e} + \sigma_e \leqslant \Sigma\}$. For any $e \in \mathbb{Z}^n$, let

$$\pi_e := \prod_{e_i \neq 0} |e_i|.$$

Given $e \in E$, consider $\pi'_e := \pi_e \prod_{e' \in E \setminus \{e\}} \pi_{e'-e}$ and note that $\pi'_e < 2^{|E|\Sigma}$. We say that $e$ is *admissible* if $e \operatorname{rem} r \neq 0$ and $(e'-e) \operatorname{rem} r \neq 0$ for all $e' \in E \setminus \{e\}$. This is the case if and only if $\pi'_e$ is not divisible by $r$. Now $\pi'_e$ is divisible by at most $\rho(2^{|E|\Sigma})$ distinct prime numbers, by Theorem 2.4. Since there are at least

$$\frac{3}{5} R / \log R \geqslant \frac{3}{5} \beta^2 S / \log(\beta^2 S)$$

prime numbers in $(R, 2R)$, by Theorem 2.3, the probability that $\pi_e$ is divisible by $r$ is at most

$$\frac{\rho(2^{|E|\Sigma})}{\frac{3}{5}\beta^2 S/\log(\beta^2 S)}.$$

From (4.1) we obtain $\Sigma \geqslant \beta \geqslant 64$, whence $2^{|E|\Sigma} \geqslant e^e$. It follows that

$$\begin{aligned}
\frac{\rho(2^{|E|\Sigma})}{\frac{3}{5}\beta^2 S/\log(\beta^2 S)} &\leqslant \frac{1.538 \log(2^{|E|\Sigma})/\log(\log(2^{|E|\Sigma}))}{0.6\,\beta^2 S/\log(\beta^2 S)} \\
&\leqslant \frac{1.538\,|E|\Sigma/\log(|E|\Sigma)}{0.6\,\beta^2 S/\log(\beta^2 S)} \\
&\leqslant \frac{1.538\,|E|\Sigma/\log(|E|\Sigma)}{0.6\,\beta^2 S/\log(\beta^2 S^2)} \\
&= \frac{1.538\,|E|\Sigma/\log(|E|\Sigma)}{0.3\,\beta^2 S/\log(\beta S)} \\
&\leqslant \frac{5.127}{\beta}\frac{\Sigma T/\log(\Sigma T)}{\beta S/\log(\beta S)} \qquad \text{(since } |E| \leqslant T\text{)}\\
&\leqslant \frac{5.127}{\beta}, \qquad\qquad\qquad\qquad\qquad (4.3)
\end{aligned}$$

since $\beta S \geqslant \Sigma T$ from (4.1). Now consider two admissible exponents $e \neq e'$ in $E$ and let $i \in \mathbb{N}_n$ with $(e_i - e'_i) \operatorname{rem} r \neq 0$. For fixed values of $\tau_j$ with $j \neq i$, there is a single choice of $\tau_i \in \{1, \ldots, r-1\}$ with $\tau \cdot (e - e') \operatorname{rem} r = 0$. Hence the probability that this happens with random $\tau_0, \ldots, \tau_{n-1}$ is $1/(r-1)$. Consequently, for fixed $e \in E$, the probability that $\tau \cdot (e - e') \operatorname{rem} r = 0$ for some $e' \in E \setminus \{e\}$ is at most

$$\frac{|E|}{r-1} \leqslant \frac{|E|}{R} \leqslant \frac{|E|}{\beta^2 S} \leqslant \frac{|E|}{\beta \Sigma T} \leqslant \frac{1}{\beta \Sigma} \leqslant \frac{1}{64\beta}, \qquad (4.4)$$

thanks to (4.1).

Assuming now that $r$ is fixed, let us examine the probability that $\bar{f}$ is $q$-faithful. Let $\pi$ be the product of all non-zero coefficients of $\bar{f}$. Then $\bar{f}$ is $q$ faithful if and only if $q$ does not divide $\pi$. Now the bit-size of the product $\pi$ is bounded by $\sigma_f \leqslant S$, by Lemma 4.1. Hence $\pi$ is divisible by at most $1.538\, S/\log S$ prime numbers, by Theorem 2.4 and our assumption that $S \geqslant 2^{16}$. With probability at least

$$1 - \frac{1}{\beta}, \qquad (4.5)$$

there are at least $R^5/(24 \log R)$ prime numbers amongst which $q$ is chosen at random, by Theorem 2.5(b). Assuming this, $\bar{f}$ is not $q$-faithful with probability at most

$$\frac{1.538\, S/\log S}{R^5/(24 \log R)} \leqslant \frac{37}{R^4} \leqslant \frac{1}{10\beta}, \qquad (4.6)$$

since $R \geqslant S \geqslant 2^{16}$, by (4.2).

Let $E'$ be the set of $e \in E$ such that $c_e \operatorname{rem} q \neq 0$ and $\tau \cdot (e - e') \operatorname{rem} r \neq 0$ for all $e' \in E \setminus \{e\}$. Altogether, the bounds (4.3), (4.4), (4.5), and (4.6) imply that the probability that a given $e \in E$ belongs to $E'$ is at least $1 - 9\varepsilon$. It follows that the expectation of $|E'|$ is at least $(1 - 9\varepsilon)|E|$. For

$$\delta := \sqrt{9\varepsilon} = \frac{3}{\sqrt{\beta}},$$

this further implies that the probability that $|E'| < (1 - 9\varepsilon/\delta)|E|$ cannot exceed $\delta$: otherwise the expectation of $|E'|$ would be $< \delta(1 - 9\varepsilon/\delta)|E| + (1-\delta)|E| = (1 - 9\varepsilon)|E|$.

We finally observe that $e \in E'$ is $\Pi_{\tau,r,q}$-faithful whenever $\tau \cdot (e-e') \operatorname{rem} r \neq 0$ for all $e' \in \operatorname{supp} f$ such that $\sigma_{c_{e'}} + \sigma_{e'} > \Sigma$. Now for every $e'$ with $\sigma_{c_{e'}} + \sigma_{e'} > \Sigma$, there is at most one $e \in E'$ with $\tau \cdot (e-e') \operatorname{rem} r = 0$: if $\tau \cdot (\tilde{e} - e') \operatorname{rem} r = 0$ for $\tilde{e} \in E'$, then $\tau \cdot (\tilde{e}-e) \operatorname{rem} r = 0$, whence $\tilde{e} = e$. By (4.1), there are at most $T/\beta$ exponents $e'$ with $\sigma_{c_{e'}} + \sigma_{e'} > \Sigma$. We conclude that $N^* \geqslant (1 - 9\varepsilon/\delta)N - T/\beta$, whenever $|E'| \geqslant (1 - 9\varepsilon/\delta)|E|$. □

## 4.3. Computing probabilistic codes for the exponents

Lemma 4.3 allows us to compute the coefficients of $\Pi_{\tau,r,q^\nu}(f(\alpha x))$ with good complexity. In the univariate case, it turns out that the exponents of $\Pi_{\tau,r,q}$-faithful terms of $f$ can be recovered as quotients of matching terms of $\Pi_{\tau,r,q^{2\nu}}((1+q^\nu)f)$ and $\Pi_{\tau,r,q^{2\nu}}(f)$ by taking $\nu$ sufficiently large.

In the multivariate case, this idea still works, for a suitable Kronecker-style choice of $\tau$. However, we only reach a suboptimal complexity when the exponents of $f$ are themselves sparse and/or of unequal magnitudes. The remedy is to generalize the "quotient trick" from the univariate case, by combining it with the algorithms from section 3: the quotients will now yield the exponents in encoded form.

Let us now specify the remaining parameters from section 3. First of all, we take $B := q^\mu$, where

$$\mu := \left\lceil \frac{6 \log S + 4 \log n + 52 \log 2}{\log q} \right\rceil.$$

Consequently,

$$2^{52} n^4 S^6 \leqslant B < 2^{52} n^4 S^6 q. \tag{4.7}$$

We also take $P := \lfloor \sqrt{B}/2 \rfloor$ and $\gamma := \lceil 6 e \log S \rceil$. Since $B$ is odd, the inequalities (3.2) hold. We compute $U$ and $\nu^{\{u\}}, p^{\{u\}}, I^{\{u\}}$ for $u = 1, \ldots, U$ as in section 3.3.

For $u = 1, \ldots, U$, $k \in \mathbb{N}_{\lambda^{\{u\}}}$, and $i \in \mathbb{N}_n$, let

$$\Pi^{\{u\}} := \Pi_{\tau,r,B^{2\nu^{\{u\}}}}$$

$$\alpha_{k,i}^{\{u\}} := \begin{cases} 1 + p_i^{\{u\}} B^{\nu^{\{u\}}} & \text{if } i \in I_k^{\{u\}} \\ 1 & \text{otherwise.} \end{cases}$$

For any term $c x^e$ with $c \in \mathbb{Z}$ and $e \in \mathbb{N}^n$, note that

$$\Pi^{\{u\}}(c x^e) = c t^{\tau \cdot e}$$
$$\Pi^{\{u\}}(c (\alpha_k^{\{u\}} x)^e) = \left(1 + \phi_k^{\{u\}}(e) B^{\nu^{\{u\}}}\right) c t^{\tau \cdot e}.$$

Whenever $c \operatorname{rem} q \neq 0$, it follows that $\phi_k^{\{u\}}(e)$ can be read off modulo $B^{\nu^{\{u\}}}$ from the quotient of $\Pi^{\{u\}}(c (\alpha_k^{\{u\}} x)^e)$ and $\Pi^{\{u\}}(c x^e)$.

LEMMA 4.5. *Let $n, S, \beta, \Sigma, R$ be as in (4.1), (4.2) and let $(r, q, \omega)$ be as in Theorem 2.5. Then we can compute the random parameters $I_k^{\{u\}}, p^{\{u\}}$ ($u = 1, \ldots, U, k \in \mathbb{N}_{\lambda^{\{u\}}}$) and $\tau_0, \ldots, \tau_{n-1}$ in time $n \tilde{O}(S)$ and with a probability of success $\geqslant 1 - 1/S$.*

**Proof.** The cost to generate $n$ random elements $\tau_0, \ldots, \tau_{n-1}$ in $\{1, \ldots, r-1\}$ is

$$O(n \sigma_r) = O(S \log S),$$

since we assumed $S \geqslant n$. The generation of $I_0^{\{u\}}, \ldots, I_{\lambda^{\{u\}}-1}^{\{u\}}$ can be done in time

$$O(\lambda^{\{u\}} m^{\{u\}} \sigma_n) = O(\gamma n \sigma_n) = \tilde{O}(S).$$

For $u=1,\ldots,U$, we compute $p^{\{u\}}$ using Lemma 2.6 with $\varepsilon := 1/(SU+1)$. The computation of a single $p^{\{u\}}$ takes time

$$O(n (\log n + \log(\varepsilon^{-1}))) (\log P)^{O(1)} = \tilde{O}(S)$$

and succeeds with probability at least $1-\varepsilon$. The probability of success for all $u=1,\ldots,U$ is at least $(1-\varepsilon)^U \geqslant 1-1/S$, because

$$\log(1-\varepsilon) \geqslant \frac{-\varepsilon}{1-\varepsilon} = \frac{-1}{SU} \geqslant \frac{\log(1-1/S)}{U}.$$

We conclude with the observation that $\log U = O(\log S)$. □

LEMMA 4.6. *Assume that $\bar{f}$ is q-faithful and that $\Pi_{\tau,r,q}(f)$ is known. Let $T' := \#\Pi_{\tau,r,q}(f) \leqslant T$. Then we can compute $\Pi^{\{u\}}(f)$ and $\Pi^{\{u\}}(f(\alpha_k^{\{u\}} x))$ for all $u \in \{1,\ldots,U\}$ and $k \in \mathbb{N}_{\lambda^{\{u\}}}$ in time*

$$\mathsf{A}_f(S) \tilde{O}(\beta \gamma S).$$

**Proof.** For a fixed $u \in \{1,\ldots,U\}$, we can compute $\Pi^{\{u\}}(f)$ and $\Pi^{\{u\}}(f(\alpha_k^{\{u\}} x))$ for all $k \in \mathbb{N}_{\lambda^{\{u\}}}$ in time

$$2\lambda^{\{u\}} \mathsf{A}_f(2\nu^{\{u\}} \mu \sigma_q) T' \nu^{\{u\}} \mu \sigma_q + \lambda^{\{u\}} n \tilde{O}((T' + \log r) \nu^{\{u\}} \mu \log q)$$

by Lemma 4.3. From Lemma 3.5 and $\mu \sigma_q = O(\log S)$, we get

$$\lambda^{\{u\}} \nu^{\{u\}} \mu \sigma_q \leqslant 18 \gamma \mu \sigma_q \Sigma = O(\gamma \Sigma \log S).$$

By definition of $R$ and $\beta$, we have $\log r = \log S + \log \beta = O(\log S)$. By (4.1) we also have $\Sigma = O(S \log^2 S)$. Hence the cost for computing $\Pi^{\{u\}}(f)$ and $\Pi^{\{u\}}(f(\alpha_k^{\{u\}} x))$ simplifies to

$$\begin{aligned}
& \mathsf{A}_f(\nu^{\{u\}} \mu \sigma_q) O(T' \gamma \Sigma \log S) + n \tilde{O}((T' + \log r) \gamma \Sigma \log S) \\
&= \mathsf{A}_f(\Sigma + \log n \log S) \tilde{O}(\gamma \beta S) + n \tilde{O}(\gamma \beta S) \quad\quad \text{(by (4.1))} \\
&= (\mathsf{A}_f(S) + n) \tilde{O}(\beta \gamma S) \\
&= \mathsf{A}_f(S) \tilde{O}(\beta \gamma S). \quad\quad\quad\quad\quad\quad\quad\quad \text{(since } n = O(\mathsf{A}_f(S)))
\end{aligned}$$

Since $U = O(\log \min(\Sigma, n)) = O(\log S)$, the total computation time for $u = 1, \ldots, U$ is also bounded by $\mathsf{A}_f(S) \tilde{O}(\beta \gamma S)$. □

Consider a family of numbers $\psi_{u,k,i} \in \mathbb{Z}_{B^{\nu^{\{u\}}}}$, where $u=1,\ldots,U$, $k \in \mathbb{N}_{\lambda^{\{u\}}}$, and $i \in \mathbb{N}_{T'}$. We say that $(\psi_{u,k,i})$ is a *faithful exponent encoding* for $f$ if we have $\psi_{u,k,i} = \phi_k^{\{u\}}(e)$ whenever $e$ is a $\Pi_{\tau,r,B^{2\nu^{\{u\}}}}$-faithful exponent of $f$ with $t^{\bar{e}_i} = \Pi_{\tau,r,B^{2\nu^{\{u\}}}}(x^e)$. We require nothing for the remaining numbers $\psi_{u,k,i}$, which should be considered as garbage.

COROLLARY 4.7. *Assume that $\bar{f}$ is q-faithful and that $\Pi_{\tau,r,q}(f)$ is known. Then we may compute a faithful exponent encoding for $f$ in time $\mathsf{A}_f(S) \tilde{O}(\beta \gamma S)$.*

**Proof.** We compute all $\Pi^{\{u\}}(f)$ and $\Pi^{\{u\}}(f(\alpha_k^{\{u\}} x))$ using Lemma 4.6. Let $c_{u,i}$ and $c_{u,k,i}$ be the coefficients of $t^{\bar{e}_i}$ in $\Pi^{\{u\}}(f)$ and $\Pi^{\{u\}}(f(\alpha_k^{\{u\}} x))$. Then we take $\psi_{u,k,i} := c_{u,k,i}/c_{u,i}$ if $c_{u,k,i}$ is divisible by $c_{u,i}$ and $\psi_{u,k,i} := 0$ if not. For a fixed $u \in \{1, \ldots, U\}$ all these divisions take $\tilde{O}(\beta \gamma S)$ bit-operations, using a similar reasoning as in the proof of Lemma 4.6. Since $U = O(\log S)$, the result follows. □

### 4.4. Sparse interpolation

Let $f = \sum_e c_e x^e$. We say that $\tilde{f} = \sum_e \tilde{c}_e x^e \in \mathbb{Z}[x_1, \ldots, x_n]$ is a $T$-approximation of $f$ if $\#\tilde{f} \leqslant \#f$ and $\#(f - \tilde{f}) \leqslant \frac{T}{2}$.

**Lemma 4.8.** *Let $(r, q, \omega)$ be as in Theorem 2.5, with $R$ as in (4.2). There is a Monte Carlo algorithm that computes a $T$-approximation of $f$ in time $\mathsf{A}_f(S)\, \tilde{O}(\beta \gamma S)$ and which succeeds with probability at least $1 - \frac{3}{\sqrt{\beta}} - \frac{17}{S}$.*

**Proof.** We first compute the required random parameters as in Lemma 4.5. This takes time $n\tilde{O}(S)$ and succeeds with probability at least $1 - 1/S$. We next compute $\Pi_{\tau,r,q}(f)$ using Lemma 4.2, which can be done in time

$$\mathsf{A}_f(\sigma_q)\, r \log q + n\tilde{O}(r \log q) = \mathsf{A}_f(S)\, O(R \log R) + n\tilde{O}(R) = (\mathsf{A}_f(S) + n)\, \tilde{O}(S).$$

We next apply Corollary 4.7 and compute a faithful exponent encoding for $f$, in time $\mathsf{A}_f(S)\, \tilde{O}(\beta \gamma S)$. By Lemma 4.4, this computation fails with probability at most $\frac{3}{\sqrt{\beta}}$. Moreover, in case of success, we have $N^* \geqslant \left(1 - \frac{3}{\sqrt{\beta}}\right) N - \frac{T}{\beta}$, still with the notation of Lemma 4.4.

From (4.1) and $S \geqslant T$, we get $T\Sigma > \beta S - T \geqslant (\beta - 1) S \geqslant S \geqslant n$. This allows us to apply Theorem 3.6 to the faithful exponent encoding for $f$. Let $\bar{e}_0, \ldots, \bar{e}_{T'-1}$ be the exponents of $\Pi_{\tau,r,q}(f)$. Given $i \in \mathbb{N}_{T'}$ such that there exists a $\Pi_{\tau,r,q}$-faithful term $cx^e$ of $f$ with $\Pi_{\tau,r,q}(x^e) = t^{\bar{e}_i}$ and $\sigma_e \leqslant \Sigma$, let us write $e_i := e$. For every such $i$, Theorem 3.6 produces a guess for $e_i$, and with probability at least

$$1 - TnU\mathrm{e}^{-\gamma/\mathrm{e}} - \frac{288\gamma U \Sigma T n^2 \log B}{P}$$

these guesses are all correct. The running time of this step is bounded by

$$\tilde{O}(\gamma T \Sigma \log B) = \tilde{O}(\beta \gamma S),$$

since $\log q = O(\log S)$, $S \geqslant n$, and (4.7) imply $\log B = O(\log S)$.

Below we will show that

$$TnU\mathrm{e}^{-\gamma/\mathrm{e}} + \frac{288\gamma U \Sigma T n^2 \log B}{P} \leqslant \frac{16}{S}. \tag{4.8}$$

Let $\nu$ be the smallest integer such that $q^\nu \geqslant 2^{\Sigma+1}$. We finally compute $\Pi_{\tau,r,q^\nu}(f)$ using Lemma 4.3, which can be done in time

$$\mathsf{A}_f(\nu \sigma_q)\, T' \nu \sigma_q + n\tilde{O}((T' + \log r)\nu \log q) = (\mathsf{A}_f(S) + n)\, \tilde{O}(\beta S) = \mathsf{A}_f(S)\, \tilde{O}(\beta S).$$

Let $c_0, \ldots, c_{T'-1} \in \mathbb{Z}_{q^\nu}$ be such that

$$\Pi_{\tau,r,q^\nu}(f) = (c_0 t^{\bar{e}_0} + \cdots + c_{T'-1} t^{\bar{e}_{T'-1}}) \bmod q^\nu.$$

For every $\Pi_{\tau,r,q}$-faithful term $c x^e$ of $f$ with $\Pi_{\tau,r,q}(x^e) = t^{\bar{e}_i}$ and $\sigma_c < \Sigma$, we have

$$\Pi_{\tau,r,q^\nu}(c x^e) = (c_i \bmod q^\nu) \, t^{\bar{e}_i},$$

so we can recover $c = c_i \in \mathbb{Z}_{q^\nu}$ from $\Pi_{\tau,r,q^\nu}(f)$.

With a probability at least $1 - \frac{3}{\sqrt{\beta}} - \frac{17}{S}$, all the above steps succeed. In that case, we clearly have $\#\tilde{f} = T' \leqslant T = \#f$. Let $f = f^{\#} + f^{\flat}$, where $f^{\#}$ (resp. $f^{\flat}$) is the sum of the terms of bit-size $>\Sigma$ (resp. $\leqslant \Sigma$), so that $N = \#f^{\flat}$. Then $\#f^{\#} \leqslant \frac{T}{\beta}$ and

$$\#(f^{\flat} - \tilde{f}) \leqslant N - N^* + \frac{T}{\beta} \leqslant \frac{3}{\sqrt{\beta}} N + \frac{2T}{\beta}.$$

Consequently,

$$\#(f - \tilde{f}) \leqslant \frac{3}{\sqrt{\beta}} N + \frac{3T}{\beta} \leqslant \frac{T}{2},$$

which means that $\tilde{f} := c_0 x^{e_0} + \cdots + c_{T'-1} x^{e_{T'-1}}$ is a $T$-approximation of $f$.

In order to conclude, it remains to prove the claimed inequality (4.8). Using the definition of $\gamma$ and the inequalities $T \leqslant S$, $n \leqslant S$, $U \leqslant \log_2 S + 2 \leqslant S$, we have

$$T n U e^{-\gamma/e} \leqslant \frac{T n U}{S^6} \leqslant \frac{1}{S^3}. \tag{4.9}$$

From (4.7) we have $B \geqslant 2^{52}$ and therefore $\sqrt{B}/(2P) \leqslant \sqrt{B}/(\sqrt{B} + 2) \leqslant 1 + 2^{-25}$. So the inequality $\Sigma T \leqslant \beta S$ yields

$$\frac{288 \gamma U \Sigma T n^2 \log B}{P} \leqslant \frac{577 \beta \gamma U n^2 S \log B}{\sqrt{B}}. \tag{4.10}$$

Let us analyze the right-hand side of (4.10). Without further mention, we will frequently use that $S \geqslant 2^{16}$. First of all, we have

$$\begin{aligned}
\sigma_S &\leqslant \log_2 S + 1 \leqslant (1/\log 2 + 1/\log(2^{16})) \log S \leqslant 1.54 \log S \\
\beta &= 64 \sigma_S^2 \leqslant 152 \log^2 S, \\
\gamma &= \lceil 6 e \log S \rceil \leqslant (16.31 + 1/\log(2^{16})) \log S \leqslant 17 \log S, \\
U &\leqslant \log_2 \Sigma + 3 \leqslant \log_2 \beta + \log_2 S + 3 \\
&= 2 \log_2 \sigma_S + \log_2 S + 9 \leqslant 2 \log_2 (\log_2 S + 1) + \log_2 S + 9 \leqslant 3 \log S.
\end{aligned}$$

It follows that

$$577 \beta \gamma U \leqslant 577 \times 152 \times 17 \times 3 \log^4 S \leqslant 2^{23} \log^4 S. \tag{4.11}$$

Now the function $x \mapsto (\log_2 x)^4 / x$ is decreasing for $x \geqslant e^4$. Consequently,

$$\frac{\sigma_S^4}{S} \leqslant \frac{2 (\log_2(2S))^4}{2S} \leqslant \frac{17^4}{2^{16}} \leqslant 2.$$

Similarly, $\sigma_S^4 / S^2 \leqslant 17^4 / 2^{32} \leqslant 2^{-15}$. Hence,

$$\begin{aligned}
R &= \max(S, 2^{58}) \beta^2 = \max(2^{12} S \sigma_S^4, 2^{70} \sigma_S^4) \leqslant 2^{55} S^2 \\
q &\leqslant R^6 \leqslant 2^{330} S^{12} \\
B &\leqslant 2^{52} n^4 S^6 q \leqslant 2^{52} S^{10} q \leqslant 2^{382} S^{22}. \tag{by 4.7}
\end{aligned}$$

This yields
$$\log B \leqslant 22 \log S + 382 \log 2 \leqslant 46 \log S. \tag{4.12}$$

Combining (4.11), (4.12), and (4.7), we deduce that
$$\frac{577 \beta \gamma U n^2 S \log B}{\sqrt{B}} \leqslant \frac{46 \times 2^{23} n^2 S \log^5 S}{\sqrt{B}} \leqslant \frac{46 \times 2^{23} n^2 S \log^5 S}{2^{26} n^2 S^3} \leqslant \frac{6 \log^5 S}{S^2}. \tag{4.13}$$

The inequalities (4.9), (4.10), and (4.13) finally yield the claimed bound:
$$T n U e^{-\gamma/e} + \frac{288 \gamma U \Sigma T n^2 \log B}{P} \leqslant \frac{1}{S^3} + \frac{6 \log^5 S}{S^2} = \frac{\frac{1}{S^2} + \frac{6 \log^5 S}{S}}{S} \leqslant \frac{16}{S}. \qquad \square$$

We are now ready to complete the proof of our main result.

**Proof of Theorem 1.1.** By definition of $R$ and thanks to Theorem 2.5, we may compute the triple $(r, q, \omega)$ in time $O((\log R)^{O(1)}) = O((\log S)^{O(1)})$, with probability of success at least $1 - \varepsilon$, where $\varepsilon := 1/\beta$.

Let $J := \lceil \log_2 T \rceil + 1$. Let $T^{\langle j \rangle} := \lceil T/2^j \rceil$ and $\Sigma^{\langle j \rangle} := \lfloor \beta S / T^{\langle j \rangle} \rfloor$ for $j = 0, \ldots, J$. Starting with $f^{\langle 0 \rangle} := 0$, we compute a sequence $f^{\langle 0 \rangle}, f^{\langle 1 \rangle}, \ldots, f^{\langle J \rangle}$ of successive approximations of $f$. Assuming that $f^{\langle j \rangle}$ is known for some $j < J$, we apply Lemma 4.8 with $f - f^{\langle j \rangle}$ and $T^{\langle j \rangle}$ in the roles of $f$ and $T$. With high probability, this yields a $T^{\langle j \rangle}$-approximation $\delta^{\langle j \rangle}$ of $f - f^{\langle j \rangle}$ and we set $f^{\langle j+1 \rangle} := f^{\langle j \rangle} + \delta^{\langle j \rangle}$.

In addition, for the evaluations of $f^{\langle j \rangle}$ on geometric sequences, we use fast multi-point evaluation. In the complexity bounds of Lemmas 4.3, 4.6, 4.8 and Corollary 4.7, one may verify that this allows us to replace $A_{f - f^{\langle j \rangle}}(s)$ by $A_f(s) + O((\log s)^{O(1)})$ for all $s \geqslant \Sigma^{\langle j \rangle}$.

The total running time is bounded by
$$J(A_f(S) + n)\tilde{O}(\beta \gamma S) = (A_f(S) + n)\tilde{O}(S).$$

Using the inequalities $J \leqslant \sigma_S + 1$ and $S \geqslant 2^{16}$, the probability of failure is bounded by
$$J\left(\frac{1}{\beta} + \frac{3}{\sqrt{\beta}} + \frac{17}{S}\right) \leqslant (\sigma_S + 1)\left(\frac{1}{2^6 \sigma_S^2} + \frac{3}{8 \sigma_S} + \frac{17}{S}\right) \leqslant \frac{1}{2}.$$

If none of the steps fail, then $\#(f - f^{\langle j+1 \rangle}) \leqslant \frac{T^{\langle j \rangle}}{2} \leqslant T^{\langle j+1 \rangle}$ for $j = 0, \ldots, J - 1$, by induction. In particular, $\#(f - f^{\langle J \rangle}) \leqslant \frac{\lceil T/2^{\lceil \log_2 T \rceil} \rceil}{2} = \frac{1}{2}$, so $f = f^{\langle J \rangle}$. $\qquad \square$

## 5. PRACTICAL VARIANTS

For practical purposes, the choice of $R \geqslant 2^{58} \beta^2$ is not realistic. The high constant $2^{58}$ is due to the fact that we rely on [43] for unconditional proofs for the existence of prime numbers with the desirable properties from Theorem 2.5. Such unconditional number theoretic proofs are typically very hard and lead to pessimistic constants. Numerical evidence shows that a much smaller constant would do in practice: see [16, section 4] and [39, section 2.2.2]. For the univariate case the complexity of the sparse interpolation algorithm in [39] is made precise in term of the logarithmic factors.

The exposition so far has also been optimized for simplicity of presentation rather than practical efficiency: some of the other constant factors might be sharpened further and some of the logarithmic factors in the complexity bounds might be removed. In practical implementations, one may simply squeeze all thresholds until the error rate becomes unacceptably high. Here one may exploit the "auto-correcting" properties of the algorithm. For instance, although the $T^{\langle j \rangle}$-approximation $\delta^{\langle j \rangle}$ from the proof of Theorem 1.1 may contain incorrect terms, most of these terms will be removed at the next iteration.

Our exposition so far has also been optimized for full generality. For applications, a high number of, say 10000, variables may be useful, but the bit-size of individual exponents rarely exceeds the machine precision. In fact, most multivariate polynomials $f$ of practical interest are of low or moderately large total degree. A lot of the technical difficulties from the previous sections disappear in that case. In this section we will describe some practical variants of our sparse interpolation algorithm, with a main focus on this special case.

## 5.1. Conducting most computations in double precision

In practice, the evaluation of our modular blackbox polynomial is typically an order of magnitude faster if the modulus fits into a double precision number (e.g. 53 bits if we rely on floating point arithmetic and 64 bits when using integer arithmetic). In this subsection, we describe some techniques that can be used to minimize the use of multiple precision arithmetic.

**Chinese remaindering.** If the individual exponents of $f$ are small, but its coefficients are allowed to be large, then it is classical to proceed in two phases. We first determine the exponents using double precision arithmetic only. Knowing these exponents, we next determine the coefficients using modular arithmetic and the Chinese remainder theorem: modulo any small prime $q$, we may efficiently compute $f \operatorname{rem} q$ using only $\#f$ evaluations of $f$ modulo $q$ on a geometric progression, followed by [24, section 5.1]. Doing this for enough small primes, we may then reconstruct the coefficients of $f$ using Chinese remaindering. Only the Chinese remaindering step involves a limited but unavoidable amount of multi-precision arithmetic. By determining only the coefficients of size $\leqslant \Sigma$, where $\Sigma$ is repeatedly doubled until we reach $S$, the whole second phase can be accomplished in time $\mathsf{A}_f(O(1)) \, O(S \log S) + O(S \log^3 S)$.

**Tangent numbers.** One important trick that was used in section 4.3 was to encode $\phi_k^{\{u\}}(e)$ inside an integer $1 + \phi_k^{\{u\}}(e) B^{\nu^{\{u\}}}$ modulo $B^{2\nu^{\{u\}}}$ of doubled precision $2\nu^{\{u\}} \sigma_B$ instead of $\nu^{\{u\}} \sigma_B$. In practice, this often leads us to exceed the machine precision. An alternative approach (which is reminiscent of the ones from [18, 30]) is to work with tangent numbers: let us now take

$$\Pi^{\{u\}} := \Pi_{\tau, r, B^{\nu^{\{u\}}}}$$
$$\alpha_{k,i}^{\{u\}} := \begin{cases} 1 + p_i^{\{u\}} \epsilon & \text{if } i \in I_k^{\{u\}} \\ 1 & \text{otherwise} \end{cases}$$

where $\Pi^{\{u\}}$ is extended to $\mathbb{Z}[x_0, \ldots, x_{n-1}][\epsilon]/(\epsilon^2)$ and where $\alpha_{k,i}^{\{u\}} \in \mathbb{Z}_{B^{\nu^{\{u\}}}}[\epsilon]/(\epsilon^2)$. Then, for any term $c x^e$, we have

$$\Pi^{\{u\}}(c x^e) = c t^{\tau \cdot e}$$
$$\Pi^{\{u\}}(c (\alpha_k^{\{u\}} x)^e) = (1 + \phi_k^{\{u\}}(e) \epsilon) c t^{\tau \cdot e},$$

so we may again obtain $\phi_k^{\{u\}}(e)$ from $\Pi^{\{u\}}(c\,(\alpha_k^{\{u\}}x)^e)$ and $\Pi^{\{u\}}(cx^e)$ using one division. Of course, this requires our ability to evaluate $f$ at elements of $((\mathbb{Z}/B^{\nu^{\{u\}}}\mathbb{Z})[\varepsilon]/(\varepsilon^2))^n$, which is indeed the case if $f$ is given by an SLP. Note that arithmetic in $(\mathbb{Z}/B^{\nu^{\{u\}}}\mathbb{Z})[\varepsilon]/(\varepsilon^2)$ is about three times as expensive as arithmetic in $\mathbb{Z}/B^{\nu^{\{u\}}}\mathbb{Z}$.

**Small prime divisors.** Although the algorithm divisors from section 2.4 is asymptotically efficient, it relies heavily on multiple precision arithmetic. If all $p_i$ and $a_k$ fit within machine numbers and $\min(n,N)$ is not too large, then we expect it to be more efficient to simply compute all remainders $a_k \operatorname{rem} p_i$. After the computation of pre-inverses for $p_i$, such remainders can be computed using only a few hardware instructions, and these computations can easily be vectorized [27]. As a consequence, we only expect the asymptotically faster algorithm divisors to become interesting for very large sizes like $\min(n,N) > 1000$. Of course, we may easily modify divisors to fall back on the naive algorithm below a certain threshold (recursive calls included); vectorization can still be beneficial even for moderate sizes [12].

**Chinese remaindering, bis.** As explained above, if $f$ has only small exponents, then multiple precision arithmetic is only needed during the Chinese remaindering step that recovers the coefficients from modular projections. If $f$ actually does contain some exponents that exceed machine precision, is it still possible to avoid most of the multiple precision arithmetic?

Let $(r,q,\omega)$ be a triple as in Theorem 2.5. In order to avoid evaluations of $f$ modulo large integers of the form $q^\nu$, we wish to use Chinese remaindering. Let $(r,q_1,\omega_1),\ldots,(r,q_\nu,\omega_\nu)$ be $\nu$ triples as in Theorem 2.5 with $q_1\cdots q_\nu > q^\nu$, the $q_i$ pairwise distinct, and where $r$ is shared by all triples. Since there are many primes $r$ for which $(2R, R^6) \cap (r\mathbb{N}+1)$ contains at least $R^5/(24\log R)$, such triples can still be found with high probability. In practice, $(2R, 10R(\log \nu)^2) \cap (r\mathbb{N}+1)$ already contains enough prime numbers.

Evaluations of $f$ modulo $q^\nu$ are now replaced by separate evaluations modulo $q_1,\ldots,q_\nu$ after which we can reconstruct evaluations modulo $q_1\cdots q_\nu$ using Chinese remaindering. However, one crucial additional idea that we used in Lemma 4.3 is that $\bar{f}$ is automatically $q^\nu$-faithful as soon as it is $q$-faithful. In the multi-modular setting, if $\bar{f}$ is $q_1$-faithful, then it is still true that the exponents of $\bar{f} \operatorname{rem} q_i$ are contained in the exponents of $\bar{f} \operatorname{rem} q_1$ for $i=1,\ldots,\nu$. This is sufficient for Lemma 4.3 to generalize.

## 5.2. The mystery exponent game

Let $d_0,\ldots,d_{n-1}$ be bounds for the degree of $f$ in $x_0,\ldots,x_{n-1}$, let $V$ be a bound for the maximal number of variables that can occur in a term of $f$, and let $p_0,\ldots,p_{n-1}$ the prime numbers from section 3.1.

Assume that our polynomial $f$ has only nonnegative "moderately large exponents" in the sense that $V \max(d_0 p_0,\ldots,d_{n-1}p_{n-1})$ fits into a machine number. Then we may simplify the setup from section 3.1 by taking

$$\begin{aligned}
m &:= \lceil n/V \rceil \\
\lambda &:= \lceil \gamma V \rceil \\
P &:= \lceil \eta n \log n \rceil \\
\phi_k(e) &:= \sum_{i \in I_k} p_i e_i \in \mathbb{N} \\
\phi(e) &:= (\phi_0(e),\ldots,\phi_{\lambda-1}(e)) \in \mathbb{N}^\lambda,
\end{aligned}$$

where $\gamma \geqslant 2$ and $\eta \geqslant 2$ are small constants and where we forget about $B$ and $\nu$. For any $\psi = (\psi_0, \ldots, \psi_{\lambda-1}) \in \mathbb{N}^\lambda$, let $\#\psi := |\{k \in \mathbb{N}_\lambda : \psi_k \neq 0\}|$. In this simplified setup, one may use the following algorithm to retrieve $e$ from $\phi(e)$:

**Algorithm 5.1**
**Input:** $\phi \in \mathbb{N}^\lambda$.
**Output:** $e$ with $\phi(e) = \psi$ or failed.

> Let $e := 0$, $\psi := \phi$, and $\mathcal{X} := \emptyset$.
> While there exist $k \in \mathbb{N}_\lambda$ and $i \in \mathbb{N}_n$ with $p_i | \psi_k \neq 0$ and $(k, i) \notin \mathcal{X}$ do:
> > Let $q := \psi_k / p_i$.
> > Let $\delta := \phi(0, \overset{(i-1)\times}{\ldots}, 0, q, 0, \ldots, 0)$.
> > If $\psi_j < \delta_j$ for some $j \in \mathbb{N}_\lambda$ or $\#(\psi - \delta) \geqslant \#\psi$, then let $\mathcal{X} := \mathcal{X} \cup \{(k, i)\}$.
> > Otherwise, update $e_i := e_i + q$, $\psi := \psi - \delta$, and $\mathcal{X} := \emptyset$.
> If $\psi = 0$, then return $e$.
> Otherwise, return failed.

The probability of success is non-trivial to analyze due to the interplay of the choices of $p_0, \ldots, p_{n-1}$ and the updates of $\psi$. For this reason, we consider the algorithm to be heuristic. Nevertheless, returned values are always correct:

PROPOSITION 5.1. *Algorithm 5.1 is correct.*

**Proof.** By construction, we have the loop invariant that $\phi(e) + \psi = \phi$, so the answer is clearly correct in case of success. The set of "problematic pairs" $\mathcal{X}$ was introduced in order to guarantee progress and avoid infinite loops. Indeed, $\#\psi$ strictly decreases at every update of $\psi$. For definiteness, we also ensured that $\psi$ remains in $\mathbb{N}^\lambda$ throughout the algorithm. (One may actually release this restriction, since incorrect decisions may be corrected later during the execution.) □

**Remark 5.2.** Algorithm 5.1 has one interesting advantage with respect to the method from section 3: the correct determination of some of the $e_i$ leads to a simplification of $\psi$, which diminishes the number of collisions (i.e. entries $\psi_k = \sum_{i \in I_k} p_i \tilde{e}_i$ such that the sum contains at least two non-zero terms). This allows the algorithm to discover more coefficients $e_j$ that might have been missed without the updates of $\psi$. As a result, the algorithm may succeed for lower values of $\gamma$ and $\lambda$, e.g. for a more compressed encoding of $e$.

**Remark 5.3.** From a complexity perspective, some adaptations are needed to make it run in quasi-linear time. Firstly, one carefully has to represent the sets $I_k$ so as to make the updates $\psi := \psi - \delta$ efficient. Secondly, from a theoretical point of view, it is better to collect pairs $(k, i)$ with $p_i | \psi_k \neq 0$ in one big pass (thereby benefiting from Lemma 2.7) and perform the updates during a second pass. However, this second "optimization" is only useful in practice when $n$ becomes very large (i.e. $n > 10000$), as explained in the previous subsection.

Algorithm 5.1 is very similar to the mystery ball game algorithm from [22]. This analogy suggests to choose the random sets $I_k$ in a slightly different way: let $\xi_0, \xi_1, \xi_2 : \mathbb{N}_n \to \mathbb{N}_\chi$ be random maps, where $\chi := \lceil \gamma n / 3 \rceil$ and $\lambda := 3\chi = 3\lceil \gamma n / 3 \rceil$, and take

$$I_{j\chi + k} := \xi_j^{-1}(\{k\}), \qquad j \in \mathbb{N}_3, k \in \mathbb{N}_\chi.$$

Assuming for simplicity that $e_i = \psi_i / p_i$ whenever $p_i | \psi_k$ in our algorithm, the probability of success was analyzed in [22]. It turns out that there is a phase change around $\gamma_{\mathrm{crit}} \approx 1.22179$ (and $\gamma_{\mathrm{crit}}/3 \approx 0.407265$). For any $\gamma > \gamma_{\mathrm{crit}}$ and $\varepsilon > 0$, numeric evidence suggests that the probability of success exceeds $1 - \varepsilon$ for sufficiently large $n$.

This should be compared with our original choice of the $I_k$, for which the mere probability that a given index $i \in \mathbb{N}_n$ belongs to none of the $I_k$ is roughly $\left(1 - \frac{1}{V}\right)^{\gamma V} \approx e^{-\gamma}$. We clearly will not be able to determine $e_i$ whenever this happens. Moreover, the probability that this does not happen for any $i \in \mathbb{N}_n$ is roughly $1 - n e^{-\gamma}$. In order to ensure that $1 - n e^{-\gamma} \geqslant 1 - \frac{1}{1000}$, say, this requires us to take $\gamma \geqslant \log n + 7$.

### 5.3. The Ben-Or–Tiwari encoding

Ben-Or and Tiwari's seminal algorithm for sparse interpolation [6] contained another way to encode multivariate exponents, based on the prime factorization of integers: given $n$ pairwise distinct prime numbers $p_0, \ldots, p_{n-1}$, we encode an exponent $e = (e_0, \ldots, e_{n-1}) \in \mathbb{N}^n$ as $\phi_{\mathrm{bt}}(e) := p_0^{e_0} \cdots p_{n-1}^{e_{n-1}}$. Ben-Or and Tiwari simply chose $p_i$ to be the $(i+1)$-th prime number. For our purposes, it is better to pick $n$ pairwise distinct small random primes $P/\log P \leqslant p_0, \ldots, p_{n-1} \leqslant P$ with $P = O(n \log n)$. Using (a further extension of) Lemma 2.7, we may efficiently bulk retrieve $e$ from $\phi_{\mathrm{bt}}(e)$ for large sets of exponents $e$.

The Ben-Or–Tiwari encoding can also be used in combination with the ideas from section 4.3. The idea is to compute both $\Pi_{\tau,r,q}(f)$ and $\Pi_{\tau,r,q}(g)$ with $g(x) := f(px) = f(p_0 x_0, \ldots, p_{n-1} x_{n-1})$. For monomials $c x^e$, we have

$$\Pi_{\tau,r,q}(c x^e) = c t^{\tau \cdot e}$$
$$\Pi_{\tau,r,q}(c (p x)^e) = c \phi(e) t^{\tau \cdot e},$$

so we can again compute $\phi_{\mathrm{bt}}(e)$ as the quotient of $\Pi_{\tau,r,q}(c (p x)^e)$ and $\Pi_{\tau,r,q}(c x^e)$.

If the total degree of $f$ is bounded by a small number $D$, then the Ben-Or–Tiwari encoding is very compact. In that case, all exponents $e$ of $f$ indeed satisfy $\phi_{\mathrm{bt}}(e) \leqslant P^D$, whence $\sigma_{\phi_{\mathrm{bt}}(e)} \leqslant D \sigma_P$. However, if $D$ is a bit larger (say $n = 100$ and $D = 10$), then $P^D$ might not fit into a machine integer and there is no obvious way to break the encoding $\phi_{\mathrm{bt}}(e)$ up in smaller parts that would fit into machine integers.

By contrast, the encoding $\phi$ from the previous subsection uses a vector of numbers that do fit into machine integers, under mild conditions. Another difference is that $\sigma_{\phi(e)} \leqslant \lceil \gamma V \rceil (\sigma_m + \sigma_D + \sigma_{2P})$ only grows linearly with $V$ instead of $D$, and only logarithmically with $D$. As soon as $n$ is large and $D$ not very small, this encoding should therefore be more efficient for practical purposes.